\documentclass[fleqn,usenatbib]{mnras}

\usepackage{newtxtext,newtxmath}

\usepackage[T1]{fontenc}

\DeclareRobustCommand{\VAN}[3]{#2}
\let\VANthebibliography\thebibliography
\def\thebibliography{\DeclareRobustCommand{\VAN}[3]{##3}\VANthebibliography}

\usepackage{graphicx}	
\usepackage{amsmath}	
\usepackage{amssymb}	

\newcommand{\eg}{e.g.,}

\newcommand{\Msun}{${\rm M}_{\odot}$}

\newcommand{\kms}{km~s$^{-1}$}

\newcommand{\Cofs}{$^{56}$Co}
\newcommand{\Nifs}{$^{56}$Ni}

\newcommand{\Mej}{$M_{\rm ej}$}
\newcommand{\KE}{$E_{\rm k}$}

\newcommand{\vph}{v_{\rm ph}}
 \def\gsim{\mathrel{\rlap{\lower 4pt \hbox{\hskip 1pt $\sim$}}\raise 1pt \hbox {$>$}}}
\def\lsim{\mathrel{\rlap{\lower 4pt \hbox{\hskip 1pt $\sim$}}\raise 1pt \hbox {$<$}}}

\title[The Interacting Supernova PTF11rka]{PTF11rka:  an interacting  supernova at the crossroads of  stripped-envelope and  H-poor super-luminous stellar core collapses\thanks{Based on observations taken at the ESO VLT under program 090.D-0440, Keck, Palomar, and Kitt Peak National Observatories.}}  

\author[E.  Pian et al.]{
E. Pian,$^{1}$\thanks{E-mail:  elena.pian@inaf.it}
P. A. Mazzali,$^{2,3}$
T. J. Moriya,$^{4,5}$
A. Rubin,$^{6}$
A. Gal-Yam,$^{7}$
I. Arcavi,$^{8,9}$
\newauthor
S. Ben-Ami,$^{10}$
N. Blagorodnova,$^{11}$
F. Bufano,$^{12}$
A. V. Filippenko,$^{13,14}$
M. Kasliwal,$^{15}$
\newauthor
S. R. Kulkarni,$^{15}$
R. Lunnan,$^{16}$
I. Manulis,$^{7}$
T. Matheson,$^{17}$
P. E. Nugent,$^{18,13}$ 
\newauthor
E. Ofek,$^{7}$
D. A. Perley,$^{2}$
S. J. Prentice,$^{19}$
O. Yaron$^{7}$
\\
$^{1}$INAF, Astrophysics and Space Science Observatory, via P. Gobetti 101, 40129 Bologna, Italy\\
$^{2}$Astrophysics Research Institute, Liverpool John Moores University, IC2, Liverpool Science Park, 146 Brownlow Hill, Liverpool L3 5RF, UK\\
$^{3}$Max-Planck-Institut f\"ur Astrophysik, Karl-Schwarzschild-Str. 1, D-85748 Garching, Germany\\
$^{4}$National Astronomical Observatory of Japan, National Institutes of Natural Sciences, 2-21-1 Osawa, Mitaka, Tokyo 181-8588, Japan \\
$^{5}$School of Physics and Astronomy, Faculty of Science, Monash University, Clayton, Victoria 3800, Australia\\
$^{6}$European Southern Observatory, Karl-Schwarzschild-Strasse 2, 85748, Garching bei M\"{u}nchen, Germany \\
$^{7}$Department of Particle Physics and Astrophysics, Weizmann Institute of Science, Rehovot 76100, Israel \\
$^{8}$School of Physics and Astronomy, Tel Aviv University, Tel Aviv 69978, Israel \\ 
$^{9}$CIFAR Azrieli Global Scholars program, CIFAR, Toronto, Canada\\
$^{10}$Harvard Smithsonian Center for Astrophysics, 60 Garden Street, Cambridge, MA 02138, USA \\
$^{11}$Department of Astrophysics/IMAPP, Radboud University, Nijmegen, The Netherlands \\
$^{12}$INAF-Osservatorio Astrofisico di Catania, Via Santa Sofia 78, I-95123 Catania, Italy \\
$^{13}$Department of Astronomy, University of California, Berkeley, CA 94720-3411, USA \\
$^{14}$Miller Senior Fellow, Miller Institute for Basic Research in Science, University of California, Berkeley, CA 94720, USA \\
$^{15}$Cahill Center for Astrophysics, California Institute of Technology, 1200 E. California Blvd. Pasadena, CA 91125, USA \\
$^{16}$The Oskar Klein Centre \& Department of Astronomy, Stockholm University, AlbaNova, SE-106 91 Stockholm, Sweden \\
$^{17}$NSF's National Optical-Infrared Astronomy Research Laboratory, Tucson, AZ, USA \\
$^{18}$Lawrence Berkeley National Laboratory, 1 Cyclotron Road, Berkeley, CA 94720, USA \\
$^{19}$School of Physics, Trinity College Dublin, The University of Dublin, Dublin 2, Ireland\\
}

\date{Accepted XXX. Received YYY; in original form ZZZ}

\pubyear{2020}

\begin{document}
\label{firstpage}
\pagerange{\pageref{firstpage}--\pageref{lastpage}}
\maketitle

\begin{abstract}
The hydrogen-poor  supernova PTF11rka ($z = 0.0744$), reported by the Palomar Transient Factory, was  observed with various telescopes starting a few days after the estimated explosion time of 2011 Dec. 5 UT and up to 432 rest-frame days thereafter.   The rising part of the light curve was monitored only in the $R_{\rm PTF}$ filter band, and maximum in this band was reached $\sim 30$ rest-frame days after the estimated explosion time.   The  light curve and spectra of  PTF11rka  are consistent with the core-collapse explosion of a $\sim$10 \Msun\ carbon-oxygen core evolved from a progenitor of  main-sequence mass 25--40 \Msun, that liberated a kinetic energy \KE $\approx 4 \times 10^{51}$\,erg,  expelled $\sim 8$ \Msun\ of ejecta, and  synthesised $\sim 0.5$ \Msun\ of \Nifs.   The photospheric spectra of PTF11rka are characterised by narrow absorption lines that point to suppression of the highest ejecta velocities ($\gsim 15,000$\,\kms).  This would be expected if the ejecta impacted a dense, clumpy circumstellar medium.  This in turn caused them to lose a fraction of their energy ($\sim 5 \times 10^{50}$\,erg), less than 2\% of which   was converted into radiation that sustained the light curve before maximum brightness.  This is reminiscent  of the superluminous SN~2007bi, the light-curve  shape and spectra of which are very similar to those of PTF11rka, although the latter is a factor of 10 less luminous and evolves faster in time.  PTF11rka  is in fact more similar to gamma-ray burst supernovae (GRB-SNe) in luminosity, although it has a lower energy and a lower \KE/\Mej\ ratio.  
\end{abstract}

\begin{keywords}
Supernova: individual: PTF11rka   -- galaxies: star formation -- stars: massive 
\end{keywords}




\section{Introduction}
\label{sec:introduction}

Hydrogen-poor  (or stripped-envelope) core-collapse supernovae \citep[SNe;][]{pianmazzali2017} represent only $\sim 30$\% of all core-collapse SNe \citep{shivvers2017,shivvers2019}, yet the fact that their inner core and ejecta are not enshrouded and screened from view by a thick H envelope, as in Type II (i.e., H-rich) SNe,  makes them more effective tracers of the explosion properties.  They exhibit a bevy of observational phenomenologies \citep{taubenberger2006,stritzi2009,benami2012,valenti2012,milisdan2013,bufano2014,liu2016,prentice2016,taddia2016,galyam2017,prentice2019,taddia2019} and their fundamental parameters (ejecta mass \Mej, kinetic energy \KE, radioactive \Nifs\ mass, progenitor mass) cover wide ranges \citep{mazzali2017,ashall2019}, which points to the diversity of their progenitors. The issue may be further compounded by the nature and composition of the circumstellar medium (CSM) and nearby environment of H-poor SNe, as clumpiness may also affect these properties \citep{shivvers2013}.  

Furthermore, despite observational differences between stripped-envelope and more massive H-poor superluminous SNe \citep[SLSNe;][]{quimby2011,chomiuk2011,inserra2013,nicholl2013,chen2015,greiner2015,nicholl2016,decia2018,kann2019,galyam2019}, connections between the two groups have been found, including intrinsic properties \citep{pastorello2010,whitesides2017},  role of the CSM \citep{shivvers2013,chen2017,jerkstrand2017,margutti2017,liu2018,lunnan2019}, and possible magneto-rotational driving  in both SLSNe and the most energetic H-poor SNe \citep{woosley2010,janka2012,metzger2015}.

Systematic searches and studies of SNe,  made possible by large area sky surveys \citep[\eg,][]{nicholl2014,lunnan2018,ho2019,moriya2019a,stritzi2020}, are sampling the full parameter space, not only by covering wide ranges of luminosities and velocities, but also by broadening the temporal extent of the investigation, with efficient and timely reactions, that unveil  the early behaviours and components, and sensitive late-epoch coverage that affords accurate  nebular-phase investigations.  This systematic approach makes unbiased detections possible and brings to evidence objects with intermediate properties that bridge seemingly separate groups.   

A case in point is the H- and He-poor (Type Ic; see, e.g., \citealt{filippenko1997} and \citealt{galyam2017} for reviews) SN PTF11rka.   First reported by  \citet{drake2012} at J2000 coordinates $\alpha = 12^{\rm h}40^{\rm m}44.84^{\rm s}$, $\delta = 12^\circ 53' 21.0''$, with a discovery date of 2012 Jan. 25 (UT dates are used throughout this paper), when it was already past its maximum brightness, PTF11rka had already been detected by the Palomar Transient Factory \citep[PTF;][]{rau2009,law2009} on 2011 Dec. 7 with the 48-inch Oschin Schmidt telescope  (P48) at Palomar Observatory.

Follow-up photometry and spectroscopy has taken place at various sites and epochs, including a late, nebular-phase observation at the ESO Very Large Telescope (VLT).  The similarity of its light-curve shape and photospheric-phase spectra with those of the superluminous, pair-instability SN candidate SN~2007bi  \citep{galyam2009,galyam2019}  made it an attractive candidate for monitoring and a possible missing link between stripped-envelope SNe (narrow-lined and broad-lined SNe~Ic) and H-poor SLSNe.   We report here on this monitoring.   A preliminary version of the pseudobolometric light curve was presented by \citet{prentice2016} who commented on its broadness.  

The redshift $z = 0.0744$, measured from our own spectra (Sections \ref{sec:specred}  and \ref{sec:obsspectra}), implies  a distance of 320 Mpc using $H_0 = 73$\,km\,s$^{-1}$\,Mpc$^{-1}$  \citep{riess2016} and a flat cosmological model with $\Omega_m = 0.31$  \citep{PlanckColl2016,wright2006}. 
Throughout the paper,  \KE/\Mej\ ratios are intended to be in units of $10^{51}$\,erg\,M$_\odot^{-1}$.


\section{Observations and Data Reduction}
\label{sec:observations}

\subsection{Optical Photometry}
\label{sec:photred}
 
Images of the PTF11rka sky area were taken with the Palomar 48-inch (P48)  telescope and the Mould-$R$ filter  \citep{ofek2012} at  many epochs prior to detection.   P48 observations with the same setup also solely covered  the rise of  the SN  flux during the first two weeks after detection.  Starting at about 20 days after detection and for about 200 days, exposures were taken also at the Palomar 60-inch telescope (P60) through $gri$ filters, while the  P48 monitoring  continued throughout.

Photometry in the fully nebular phase ($\sim 430$ rest-frame days after explosion) was obtained at the VLT with the FOcal Reducer and low dispersion Spectrograph 2 \citep[FORS2;][]{appenzeller1998}  through $BVRI$ filters.   
Typical exposure times were 60--180\,s at the P48 and P60 and 5--10\,min at the VLT. The seeing was on average $\sim 2^{\prime\prime}$.

For image reductions we followed \citet{laher2014}.   A high-quality image produced by  stacking several frames of the same field obtained prior to the explosion was used as a background reference.  After debiasing and flat-fielding,  the images were background-subtracted and calibrated against stars catalogued in the Sloan Digital Sky Survey (SDSS),  including both colour and colour-airmass terms to determine the zero point  of each image.  The SN magnitudes were derived using point-spread-function (PSF) fitting photometry \citep{sullivan2006,firth2015}.  The log of photometric observations  is reported in Table \ref{tab:logphotomobs}.

\subsection{Optical Spectroscopy}
\label{sec:specred}

Six low-resolution optical spectra were acquired with various telescopes and setups (see Table \ref{tab:logspecobs} for a log of the spectroscopic observations). 
The Keck spectra were obtained with the Low-Resolution Imaging Spectrometer (LRIS; \citealp{oke1995}); an atmospheric dispersion corrector was used to ensure accurate relative spectrophotometry \citep{filippenko1982}.  The  spectral frames were wavelength calibrated using standard lamp spectra and fully  reduced following the LRIS dedicated pipeline \citep{perley2019}.   
A similar observing setup and reduction method were adopted for the KPNO spectrum.
Eight exposures of 30 minutes each were acquired with the VLT and FORS2 between 2011 Mar. 11 and 15, with  the slit oriented along the parallactic angle, to minimize the effects of atmospheric dispersion.    
The  reduction of these individual spectra was carried out using IRAF and IDL routines and the spectra were finally coadded.
Flux  calibrations were applied using a solution derived from observations of standard  stars, and then adjusted  against the simultaneous photometry.


\section{Results}
\label{sec:results}

\subsection{Photometry}
\label{sec:obsphotom}

\begin{figure*}
\includegraphics[width=\textwidth]{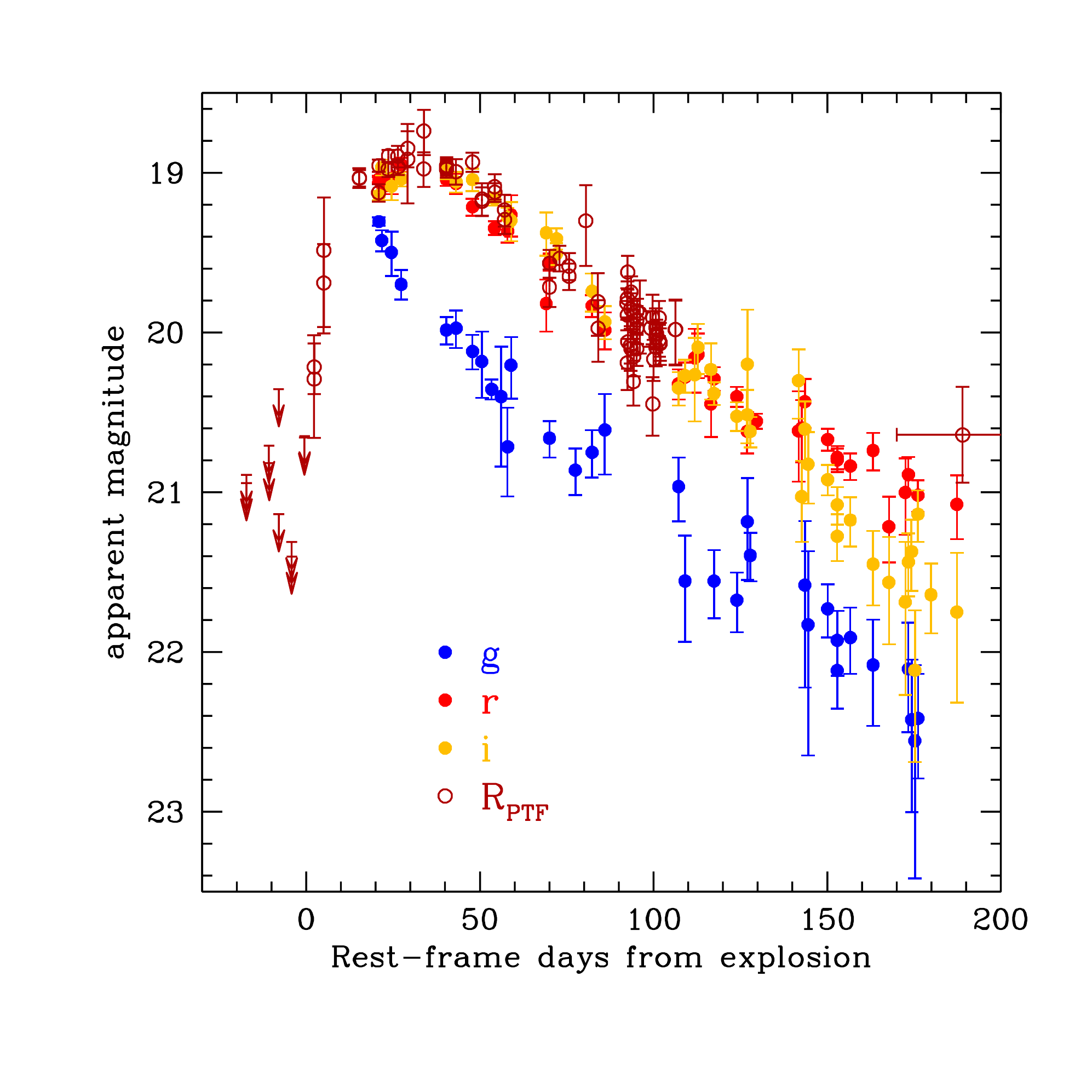}
\caption{Multiband light curves of PTF11rka in the rest frame (the photometry at $t = 430$ days is not included).  The zero point of the abscissa corresponds to the estimated explosion time (2011 Dec. 5).  The magnitudes are corrected for Galactic extinction ($E(B-V) = 0.034$ mag), K-corrected,  and host-galaxy-subtracted using a  template derived from the VLT spectrum (see Section \ref{sec:obsspectra} and Fig. \ref{fig:vltnebspec}).  The $R_{\rm PTF}$ data points between days 170 and 200 have been averaged.} 
\label{fig:mwllcs}
\end{figure*}

PTF11rka was  first detected in the $R_{\rm PTF}$ band on 2011 Dec. 7 while rising in flux (Table \ref{tab:logphotomobs} and Fig. \ref{fig:mwllcs}), with the latest nondetection dating just 3 days earlier.  We assumed the SN to have exploded about half a day after the last upper limit -- that is, around 2011 Dec. 5 (MJD = 55,900), and this is our adopted explosion date to which we have referred all photometric and spectral phases. The model light curve (see Section \ref{sec:lcmodel}) supports this assumption.

In Figure \ref{fig:mwllcs} we report the light curves in all filters up to day $\sim 200$ after explosion.  A correction for Milky Way dust absorption, K-correction (using our own spectra, see Section \ref{sec:obsspectra}, and interpolating the corrections at the photometry epochs) and host-galaxy contribution (estimated from the nebular spectrum, see Section \ref{sec:obsspectra}) were applied to the data.     For the Galactic extinction we used $A_V = 0.094$ mag  \citep{schlafly2011} and adopted the extinction curve of  \citet{cardelli1989}, with $R_V = 3.08$.  Since our host-galaxy model has very little intrinsic extinction and there is no independent evidence that the supernova is absorbed in the rest frame, we have not evaluated this correction. The $R_{\rm PTF}$ data points between  rest-frame days 170 and 200 are rather noisy and were averaged in Figure \ref{fig:mwllcs}, although the individual measurements are reported in Table \ref{tab:logphotomobs}.   
Maximum  light in the $g$ band is not covered, and it must occur at least 15 rest-frame days earlier than in the $r$ and $i$ bands, indicating strong chromatic  dependence of the SN time behaviour.

\begin{figure*}
\includegraphics[width=\textwidth]{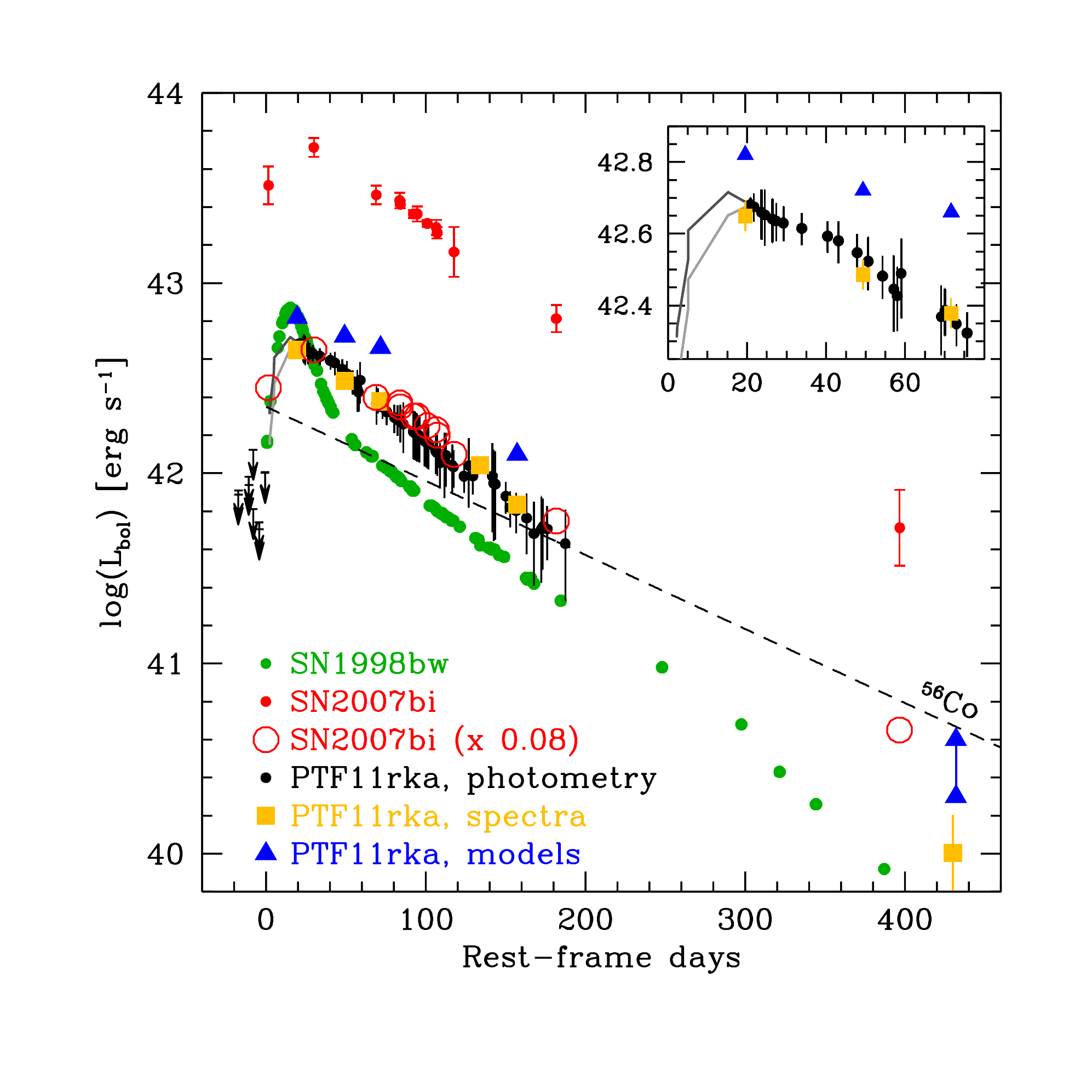}
\caption{Pseudobolometric light curve of PTF11rka (black), compared with those of H-poor energetic SN 1998bw (green) and superluminous SN~2007bi (red)  in the  rest-frame 3500--9500\,\AA\  range.  The early portion of the light curve of PTF11rka (see inset), which is based only on $R_{\rm PTF}$ measurements, is represented as curves computed under the assumption of constant $g - r$ colour (light grey) and variable  $g - r$ colour (dark grey).  The  upper limits were obtained from the $R_{\rm PTF}$ upper limits under the assumption of  a constant bolometric correction.   The yellow squares represent the  fluxes from the spectra of PTF11rka, integrated over the same wavelength range.   The uncertainties of the SN 1998bw points are equal to, or smaller than, the symbol size. The zero of time  corresponds to the explosion epoch for PTF11rka and SN 1998bw.    Note that the pseudobolometric light-curve maximum of PTF11rka seems to precede the $R_{\rm PTF}$-band maximum by $\sim 10$ days (Fig. \ref{fig:mwllcs}), but this is highly uncertain (see Section \ref{sec:obsphotom}).    The  phases of SN~2007bi, whose explosion time is very poorly determined, were shifted so that its time of maximum brightness matches the $R_{\rm PTF}$-band maximum time of PTF11rka (30 days after explosion).   The first point  of the light curve of SN~2007bi was obtained  from the $r$-band light curve reported by Gal-Yam et al. (2009)  by assuming a difference between $r$-band and pseudobolometric flux equal to that at peak luminosity (the  uncertainty is reflected in the relatively large error bar). The  large red open circles  represent the pseudobolometric light curve of SN~2007bi scaled down to match the luminosity of PTF11rka.  The two light curves are very similar. The blue triangles are the input luminosities of the spectral models  (see Section \ref{sec:specmodels}); for the nebular epoch (day 432), both low- and high-mass solutions are shown.   The  $^{56}$Co radioactive  decay law  is shown for reference (dashed line).  For all SNe  we adopted the concordance cosmology and $H_0 = 73$\,km\,s$^{-1}$\,Mpc$^{-1}$ (see Section \ref{sec:introduction}).}  
\label{fig:3bollc}
\end{figure*}

From the multiband  photometry we have evaluated a pseudobolometric light curve in the rest-frame wavelength interval  3500--9500\,\AA. We first constructed broad-band energy distributions spaced by 1 day from the interpolated $g,r,i$ and $R_{\rm  PTF}$ light curves; then we integrated the flux over the wavelength ranges of filter sensitivities and have extrapolated the flux to 3500\,\AA\ and 9500\,\AA\ assuming a flat spectrum redward and blueward of the available filter ranges.  The resulting light curve was finally remapped to the epochs of actual observations and is reported in Figure \ref{fig:3bollc}.  The pseudobolometric flux at the epoch of the last photometric measurement (rest-frame day 430 after explosion), which is dominated by line emission,  was estimated by integrating the simultaneous flux-calibrated and corrected nebular spectrum (see Section \ref{sec:obsspectra}).

Since the wavelength interval for pseudobolometric integration  (3500--9500\,\AA)  is significantly wider than the combined range of the sensitivities of the $g$, $r$, $i$, and $R_{\rm PTF}$ filters, we have verified that the pseudobolometric luminosity of PTF11rka estimated over this range is consistent with that derived from the spectra once corrected for extinction, host-galaxy contribution, and redshift.  We have integrated the spectral flux in the first five spectra  over 3500--9500\,\AA, converted it to luminosity, and reported it in Figure \ref{fig:3bollc}. This comparison shows good consistency, indicating that our method for evaluating the pseudobolometric light curve from the photometry is reliable.  

At epochs prior to maximum brightness, only the P48 photometry in the $R_{\rm PTF}$ filter is available, making the pseudobolometric flux estimate uncertain.  Therefore, we have  computed the pseudobolometric magnitudes at these epochs following two methods. First, we assumed a constant correction equivalent to the difference between the $R_{\rm PTF}$  magnitude and the pseudobolometric magnitude at maximum brightness. Second, we assumed that the $g - r$ colour changes from $-0.3$ mag at about 20 days before maximum light  \citep{prentice2016}  to the observed value $g - r = 0.25$ mag at the epoch of the first $g$-band observation, while the  $r - i$ colour stays constant (as is the case past maximum).  The simulated curves  obtained based on these two assumptions are reported in  Figure \ref{fig:3bollc}.  The absence of pre-maximum data in other bands than the $R_{\rm PTF}$ filter (the $g$-band data only cover the post-maximum decline, Fig. \ref{fig:mwllcs}) prevents an accurate estimate of the epoch of pseudobolometric maximum, which therefore must be regarded as extremely uncertain (see also Section \ref{sec:lcmodel}).

In Figure \ref{fig:3bollc} we compare the pseudobolometric light curve of PTF11rka with those of the GRB-SN 1998bw \citep{galama1998,patat2001} and the SLSN~2007bi \citep{galyam2009,young2010}, constructed from the original photometry in wavelength ranges close to that  adopted for PTF11rka and following a procedure similar to the one described above.  For SN2007bi,   \citet{young2010} estimate a pseudobolometric light curve that lacks, however, information on the rising phase.  Since \citet{galyam2009} report a few early, pre-maximum points in the $r$ band,  we have evaluated a pseudobolometric point from these $r$-band measurements assuming their difference with respect to the pseudobolometric flux is the same as at maximum light \citep[Fig. \ref{fig:3bollc}; see also][]{moriya2010}.    The sparseness of the early coverage causes  the maximum of the  light curve of SN~2007bi to be also  very poorly constrained in time.   The same cosmological model adopted in this paper (Section \ref{sec:introduction}) was applied to all light curves. 

\subsection{Spectroscopy}
\label{sec:obsspectra}

\begin{figure*}
\includegraphics[width=\textwidth]{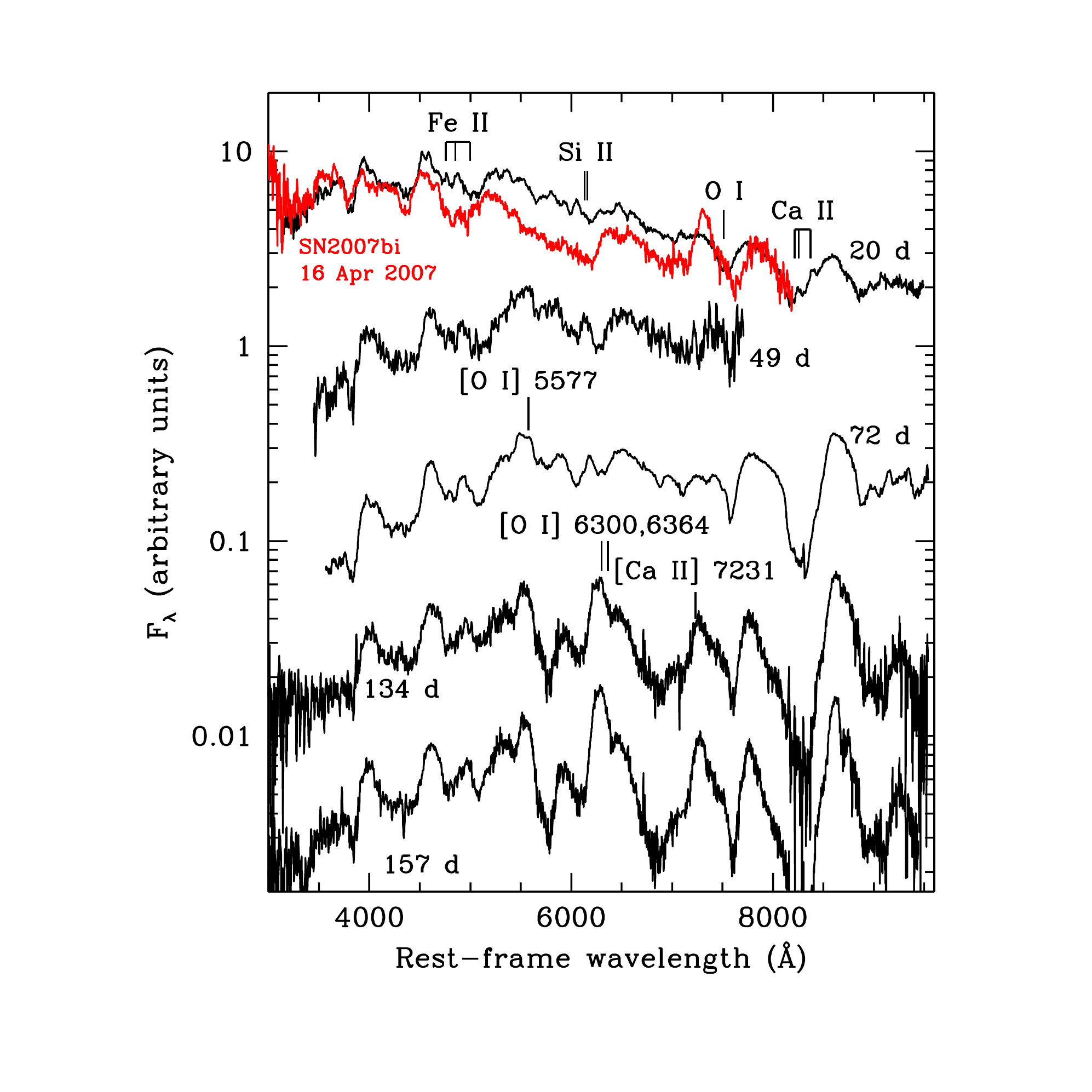}
\caption{First five spectra of PTF11rka, corrected for Galactic extinction ($E(B-V) = 0.034$ mag),  redshift ($z = 0.0744$),  and host galaxy contribution (evaluated as detailed in  Section \ref{sec:obsspectra}).  The phase is indicated close to each spectrum as rest-frame days with respect to explosion.  The absorption and emission features that are discussed in Section \ref{sec:specmodels} are indicated with vertical bars.  The absorption line marks are placed at the wavelengths  expected for a  blue-shift caused by an expansion velocity of 10,000 \kms. For comparison,  the first recorded spectrum  of SN~2007bi (corrected for $z = 0.1279$ and extinction $E(B-V) = 0.024$ mag) is shown (red) at a rest-frame phase that could be $\sim 50$ rest-frame days after maximum light or $\sim 120$ rest-frame days after explosion.    All spectra are smoothed with a 10\,\AA\ boxcar and arbitrarily scaled in flux density.} 
\label{fig:specsequence}
\end{figure*}

The spectra were first inspected for emission lines from the host galaxy; the most prominent among these is H$\alpha$, whereby $z = 0.0744$ was measured.   Before further analysis and modelling, emission lines and spurious features were discarded.   The first five spectra (see Table \ref{tab:logspecobs}), dereddened, deredshifted, and galaxy-subtracted (see below for construction of the spectral template), are reported in Figure \ref{fig:specsequence}.   They  lack H and He absorption lines, supporting a classification of PTF11rka as a SN~Ic.  According to the empirical classification scheme developed by \citet{prentice2017}, based on the number of absorption features seen in the optical spectra of He-poor SNe,  which is a measure of the degree of line blending, PTF11rka belongs to the SN~Ic-7 group (see Section \ref{sec:photspecmodel20d}). The lines are not too blended and are relatively narrow  ($\lesssim 12,000$\,\kms), pointing to non-extreme photospheric velocities and a low \KE/\Mej\  ratio  ($\lsim 1$).

Figure \ref{fig:specsequence} also reports the first spectrum acquired of SN~2007bi, at an epoch that may have been $\sim 50$ rest-frame days past peak brightness and $\sim 120$ rest-frame days past explosion. Caution is in order in the comparison with PTF11rka, as both the explosion and maximum-light epochs of SN~2007bi are very poorly determined \citep{galyam2009,young2010}.    The remarkable similarity  suggests a faster spectral evolution of PTF11rka.  Close to light-curve peak, the photospheric velocity of PTF11rka must be similar to that measured for SN~2007bi at $\sim 50$ days after peak, $\sim 12,000$\,\kms.

\begin{figure*} 
\includegraphics[width=\textwidth]{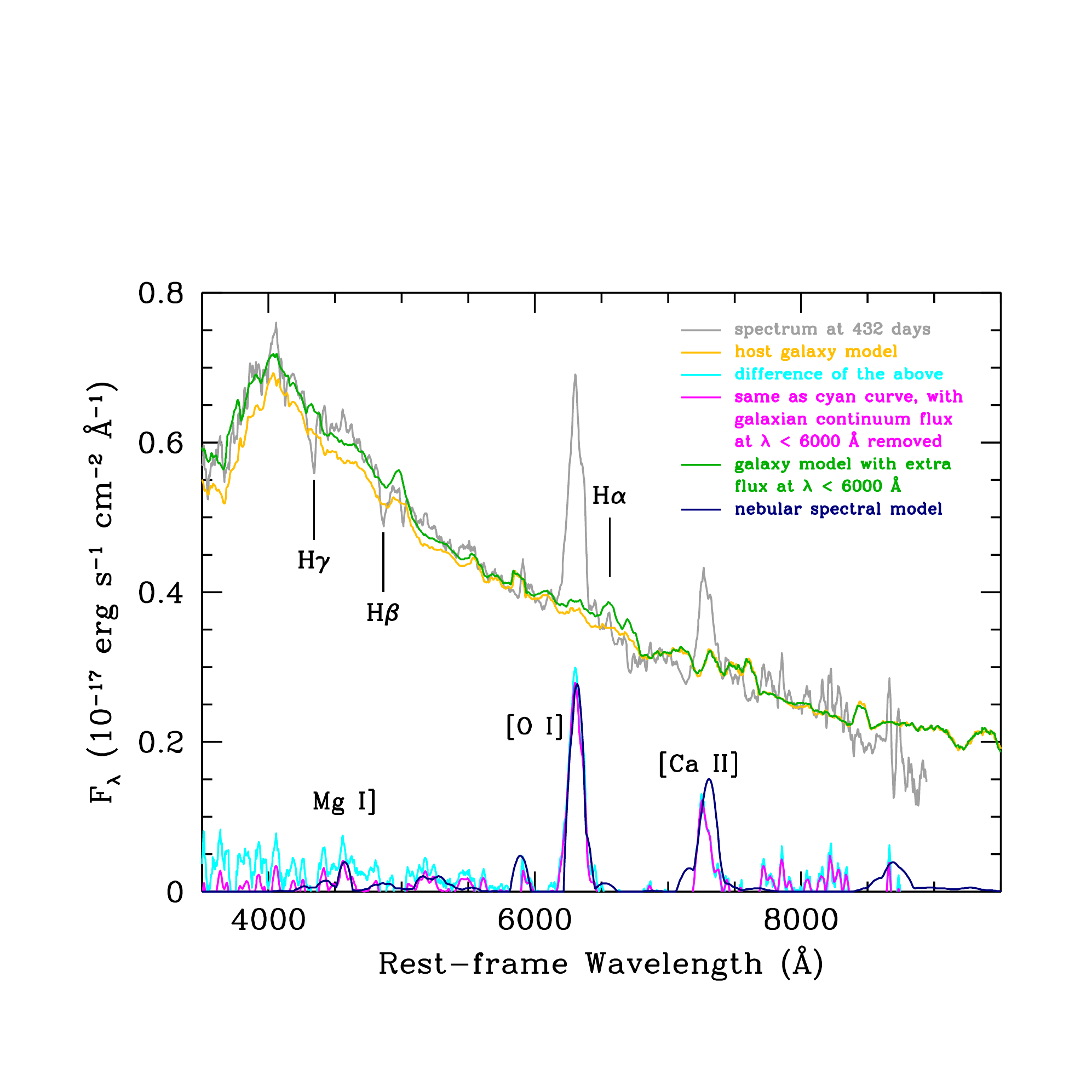}
\caption{Spectrum (grey)  taken on 2013 Mar. 13 (432  rest-frame days  after the explosion),  flux-calibrated against the simultaneous photometry, corrected for Galactic extinction ($E(B-V) = 0.034$\,mag), de-redshifted, and smoothed with a  10\,\AA\ boxcar.  The adopted host-galaxy template, from  \citet{kinney1996},  is shown in yellow and the difference between the two spectra in cyan.  A small residual continuum contribution (perhaps due to the presence of a star-forming region underlying the SN)  is still present shortward of 6000\,\AA\  and further subtracted to obtain the final  decomposed spectrum (magenta).  This extra continuum is added to the galaxy template to yield the actual galaxy flux at the SN location (green).  The nebular spectral model  presented in Section \ref{sec:nebspecmodel} is shown in dark blue and the strongest emission lines are marked.}
\label{fig:vltnebspec}
\end{figure*}

Host-galaxy stellar light dominates the continuum  of the spectrum taken 432 rest-frame days after explosion, as revealed by its shape (Fig.  \ref{fig:vltnebspec}).  This is well described with a template of  a star-forming galaxy with minimal intrinsic extinction \citep{kinney1996}.   After appropriate flux normalisation, we subtracted  this galaxy template from the spectrum.   The residual spectrum  still contains some very low-level flux at blue wavelengths (cyan curve in Figure  \ref{fig:vltnebspec}), which we attribute to additional stellar light not taken into account by the star-forming galaxy template of \citet{kinney1996}.  We therefore removed this extra blue continuum and added it to the  galaxy spectrum, to obtain a realistic template (green curve) that we adopted for subtraction from all previous spectra and from the photometry (see Section \ref{sec:obsphotom}).   Note  that  the spectral energy distribution of the galaxy constructed from SDSS $ugriz$ magnitudes\footnote{$u = 21.83$,  $g = 21.33$, $r = 21.22$,  $i = 20.9$, $z =  20.76$ mag; skyserver.sdss.org/dr13} is consistent with the galaxy spectral template after downscaling by a factor of $\sim 3$, accounting for the fact that only part of the galaxy flux was included in the VLT/FORS2 slit.

The spectral signal remaining after decomposition of the  late-epoch spectrum (Fig.\ref{fig:vltnebspec})  is  only due to  forbidden emission lines,  like [O~I] $\lambda\lambda6300$, 6364, and lower-intensity [Fe~II] $\lambda$5200 and calcium emission, that are commonly seen during the nebular phase of core-collapse SNe  \citep[see, e.g.,][]{mazzali2007a,mazzali2007b}.  There are several differences in the nebular spectrum of PTF11rka  with respect to that of SN~2007bi  \citep{galyam2009}. Notably, the weakness of the iron lines  indicates that   PTF11rka  synthesised  much less \Nifs\  (see Section \ref{sec:nebspecmodel}),  as is independently inferred from the light curve (Section \ref{sec:lcmodel}).


\section{Models}
\label{sec:models}

The unusual light curve of PTF11rka (which was almost as luminous as a GRB-SN; see Fig. \ref{fig:3bollc}) and its early-time spectra, characterised by rather sharp, narrow, unblended absorption lines,  are reminiscent of  SN~2007bi, which was thought to be a pair-instability SN candidate \citep{galyam2009}, although this was later challenged \citep{jerkstrand2016,moriya2019b,mazzali2019}.  These features 
were reproduced by \citet{moriya2019b} by adopting a sharp cut in the ejecta distribution with velocity, at 13,000\,\kms.   Physically, this may indicate that the fastest, outermost ejecta  have been slowed down by the impact on a stationary or slowly moving CSM. In deriving the properties of PTF11rka via radiation transport modelling we have assumed a similar scenario.  The CSM is likely clumpy, as it must allow SN   radiation to come through, and  unlikely to contain  H, as no H$\alpha$ emission line is seen.  This is plausible if it was material lost from the inner layers of a stripped progenitor \citep[one such example is the ``super-Chandra'' SN~Ia candidate SN 2009dc; ][]{hachinger2012}. 

In our spectra of PTF11rka  (Table \ref{tab:logspecobs} and Fig. \ref{fig:specsequence}) we see a gradual transition from photospheric to nebular conditions, with the three spectra near and just after maximum light (20, 49, and 72 rest-frame days after explosion)  showing little evolution in properties, and of the two late-epoch spectra  (157 and 432 rest-frame days after explosion) only the latter being fully nebular.   
Typically, nebular spectra can be used to quantify the amount of mass ejected by the SN, and to assess the contribution of radioactive powering. However, our fully nebular spectrum has very low signal-to-noise ratio  except in a few emission lines and does not easily lend itself to interpretation.
Therefore, our strategy is to use the earliest spectrum (20\,d), which was taken at an epoch that may have been close to pseudobolometric maximum, to determine the properties of the outer layers (\eg\ velocity cut, \KE), and the light curve to estimate the amount of mass ejected, \Mej. Both the light curve and the late-time spectrum are then used to estimate the mass of \Nifs\ synthesised in the explosion.   We have not modelled the spectrum at 134 days, as it is very similar to the one at 157 days (Fig. \ref{fig:specsequence}).

\subsection{Spectra}
\label{sec:specmodels}

We modelled the photospheric-phase spectrum with the code 
developed by \citet{mazzalilucy1993,lucy1999,mazzali2000a}, which uses the
Monte Carlo approach  for spectrosynthesis in expanding SN ejecta.
First we selected an explosion model. The luminosity and persistence of the light curve suggest rather high \Mej\ and \Nifs\ masses. The narrow absorption lines suggest that \KE\ is not too high, particularly in relation to \Mej. This, however, does not mean that significant \KE\ could not be dissipated  upon impact with CSM.  How much this impact reflects on the light curve is another question that we try to address (Section \ref{sec:lcmodel}). 

The model elaborated by \citet{moriya2010} for the light curve of SN~2007bi required a  \Nifs\ mass, ejecta mass, and kinetic energy of  6\,\Msun, 40\,\Msun,  and $3.6\times 10^{52}$\,erg, respectively.   
We adopted here the same homologous density structure of that ejecta model and scaled it to match the properties of PTF11rka. For a given ejecta mass \Mej\  and kinetic energy \KE,  the velocity $v$ and density $\rho$ scale as $v \propto$ (\KE/\Mej)$^{0.5}$  and $\rho \propto$ (\Mej$^5$/\KE$^3$)$^{0.5}$.   We rescale the model to achieve \Mej\ $\approx 8$\,\Msun\ and \KE\ $\approx 4 \times 10^{51}$\,erg. As in the case of SN~2007bi, these values yield a low ratio \KE/\Mej\ $\approx 0.5$. 
We then proceeded to model the spectra, and modified the density and abundance structure as best suited to fit the observations. 

For the nebular regime, we computed synthetic  spectra using our nonlocal thermodynamic equilibrium
code  \citep[\eg,][]{mazzali2007a}. The code first computes the deposition of the gamma rays and positrons produced in the decay of \Nifs\ and \Cofs\ in the expanding SN nebula, as described by \citet{cappellaro1997,mazzali2001a}. The energy deposited by the gamma rays and positrons is turned into heating of the gas, as described by \citet{axelrod1980}. Fast particles produced in the decay are responsible for ionisation, while recombination depends on density. The heating is then balanced by cooling, which takes place via emission mostly in forbidden lines, although some allowed transitions also contribute. While the code assumes microscopic abundance mixing, it can deal with density and abundance distributions and with clumping. Density and abundance distributions are useful to describe more accurately the emission-line profiles, or can be used when a specific explosion model is adopted. Clumping turns out to be a requirement for SN~Ic spectra. Only significant clumping (volume filling factor $\zeta \approx 0.1$) suppresses doubly ionised species and allows all cooling to take place in singly ionised species such as those observed in SNe~Ib/c at late times \citep[\eg][]{mazzali2001a}.

\subsubsection{Spectrum at phase 20 d}
\label{sec:photspecmodel20d}

For the outer ejecta layers, we use a composition that is dominated by C and O, in keeping with the properties of the outer CO core of a massive star. We use a small Si abundance (0.4\%) in view of the weak observed Si~II $\lambda\lambda6347$, 6371 line.

\begin{figure*} 
\includegraphics[width=0.8\textwidth]{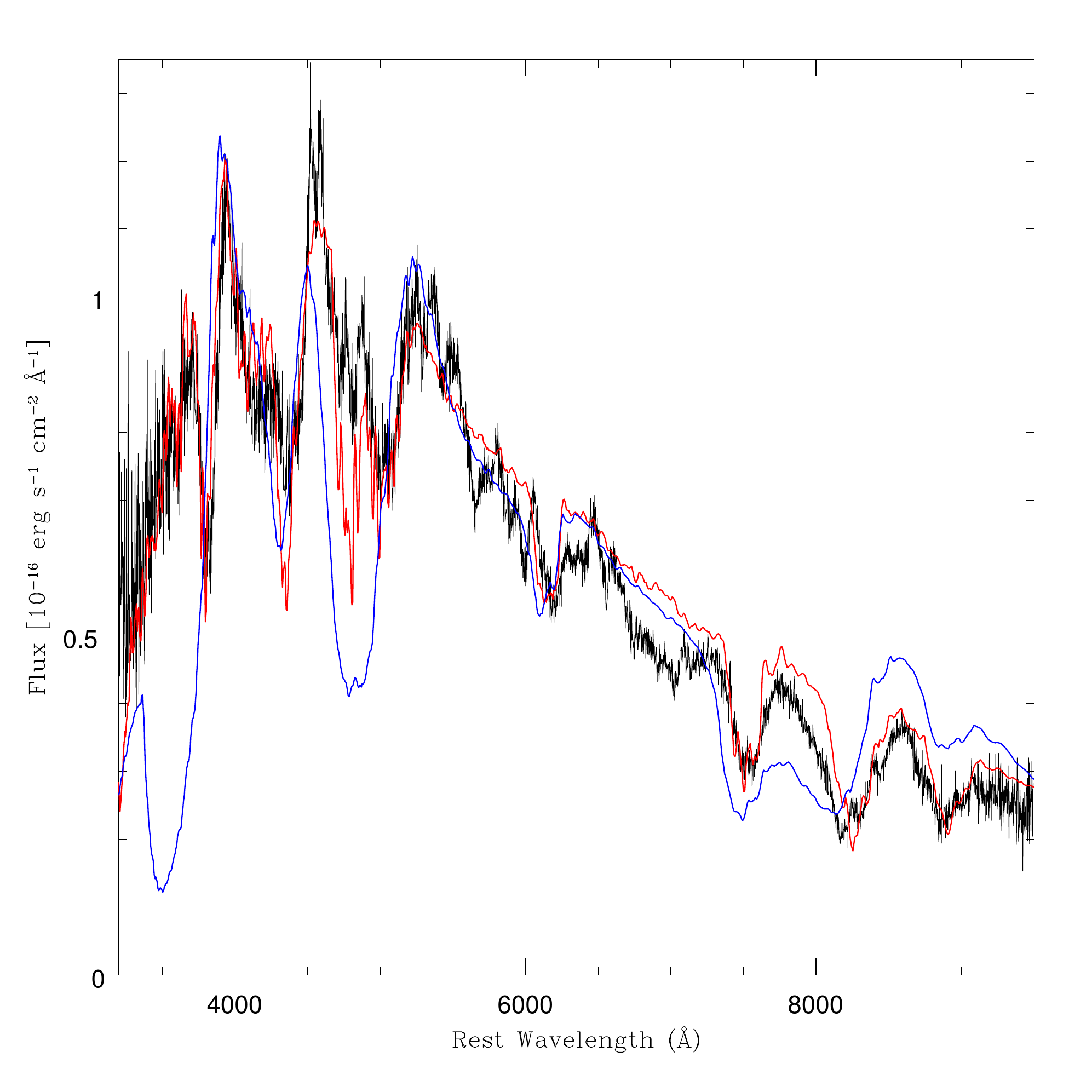}
\caption{Spectrum (black) taken on 2011 Dec. 26  (20 rest-frame days after explosion), flux-calibrated and  corrected as in Fig. \ref{fig:specsequence}.   The model shown as a blue curve  was obtained by applying no cut to the density  profile.  The red curve represents the model where the ejecta density distribution profile is cut at   a velocity  of $15,000$\,\kms.}
\label{fig:26Dec_cutnocut}
\end{figure*}

For an input luminosity  $\log L = 42.78$\,[erg\,s$^{-1}$] and photospheric velocity $\vph = 13,000$\,\kms\ we obtain a reasonable fit to the overall spectral distribution  (Fig. \ref{fig:26Dec_cutnocut}, blue curve), although all lines are too broad.   In the synthetic spectrum, the  near-infrared  Ca~II  triplet blends with O~I $\lambda$7774, which is typical of energetic ejecta, and so do the lines of Fe~II  multiplet 42 at 4800--5000\,\AA. The synthetic spectrum could be classified as a SN~Ic-3.5 in the classification of \citet[][or SN~Ic-4.5 if we included an unseen Na~I~D line]{prentice2017}. However, in the observed spectrum these lines are unblended: it shows distinct near-infrared   Ca~II  triplet  and O~I $\lambda$7774 features, making PTF11rka a SN~Ic-6 (7) in that classification. 

As it was done for SN~2007bi in \citet{moriya2019b}, we then proceeded to ``cut'' the ejecta distribution at an outer velocity of 15,000\,\kms. As shown in Fig. \ref{fig:26Dec_cutnocut} (red curve), this reproduces the desired spectral absorption features (although the Fe~II lines may be too deep). With the outer ejecta removed, the explosion model now has \Mej\ $\approx 7.9$\,\Msun\  and \KE\ $\approx 3.5 \times 10^{51}$\,erg. A best fit requires $\log L = 42.82$\,[erg\,s$^{-1}$] and $\vph = 12,500$\,\kms, and a similar composition as in the model without the cut in density.  Therefore, some $0.1$\,\Msun\ of ejecta, carrying $\sim 5 \times 10^{50}$\,erg of \KE, have been used in the collision with a CSM. The collision may have produced radiation, which supported the light curve, especially at early times, so it is possible that our estimated luminosity is too high. In any case, we used the modified model of the ejecta in the rest of the simulation. 

\subsubsection{Spectrum at phase 49 d}
\label{sec:photspecmodel49d}

\begin{figure*} 
\includegraphics[width=0.8\textwidth]{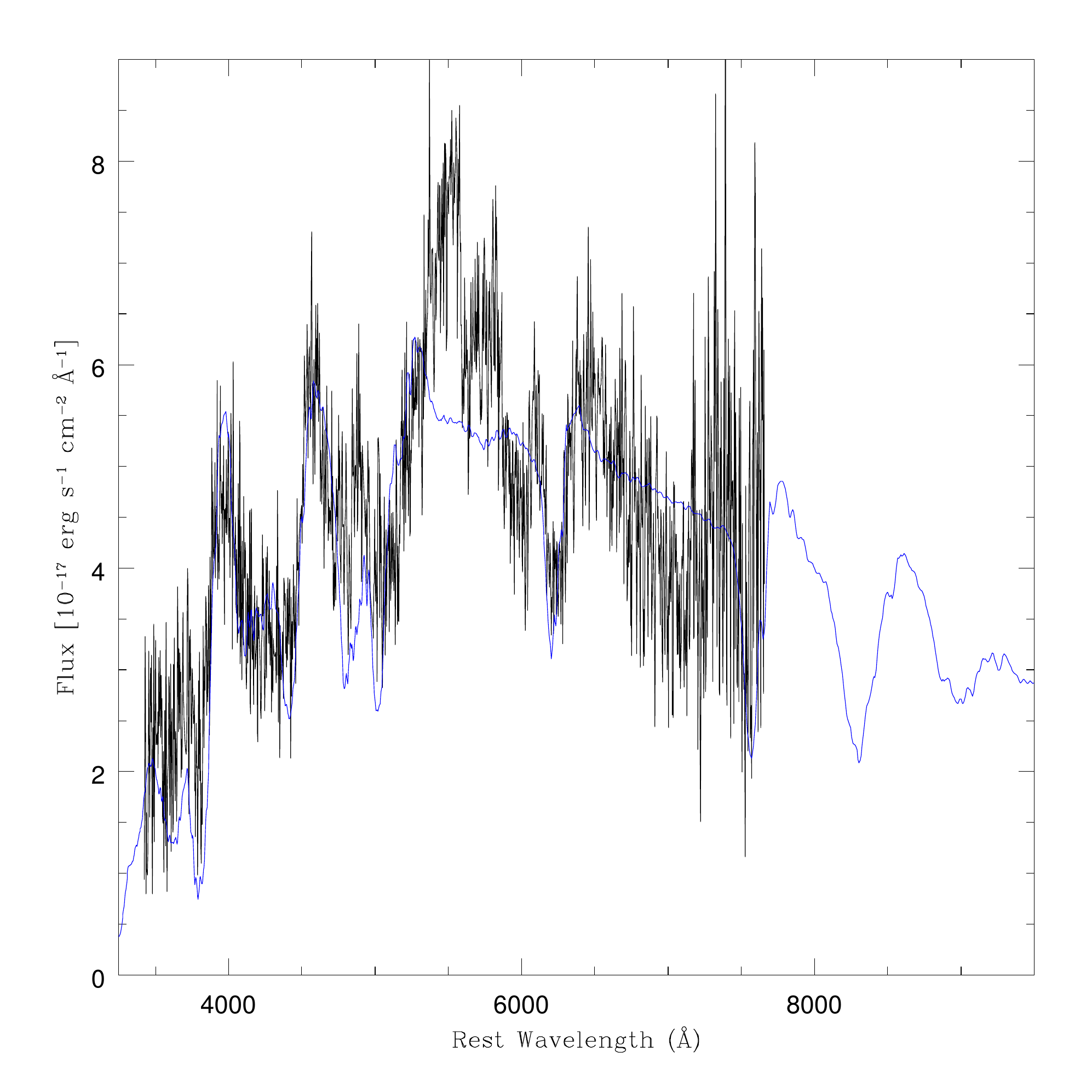}
\caption{Spectrum (black)  taken on 2012 Jan. 27  (49 rest-frame days after the explosion), 
flux-calibrated and  corrected as in Fig. \ref{fig:specsequence}.   The model shown as a blue curve  was obtained by applying a velocity cut to the ejecta density  profile at 15,000 \kms.}
\label{fig:27Jan}
\end{figure*}

The next spectrum (49 rest-frame days past explosion)  is much redder than the previous one.  The luminosity decreases to $\log L = 42.72$\,[erg\,s$^{-1}$], while the photospheric velocity decreases to 8000\,\kms, consistent with the long time elapsed since the earlier spectrum. The data at this epoch are very noisy, but the fit looks reasonable (Fig.\,\ref{fig:27Jan}). The near-photospheric composition changed somewhat, C being replaced largely by O and Si, as is expected in deeper stellar layers (some 2\,\Msun\ are now above the photosphere).

\subsubsection{Spectrum at phase 72d}
\label{sec:photspecmodel72d}

\begin{figure*} 
\includegraphics[width=0.8\textwidth]{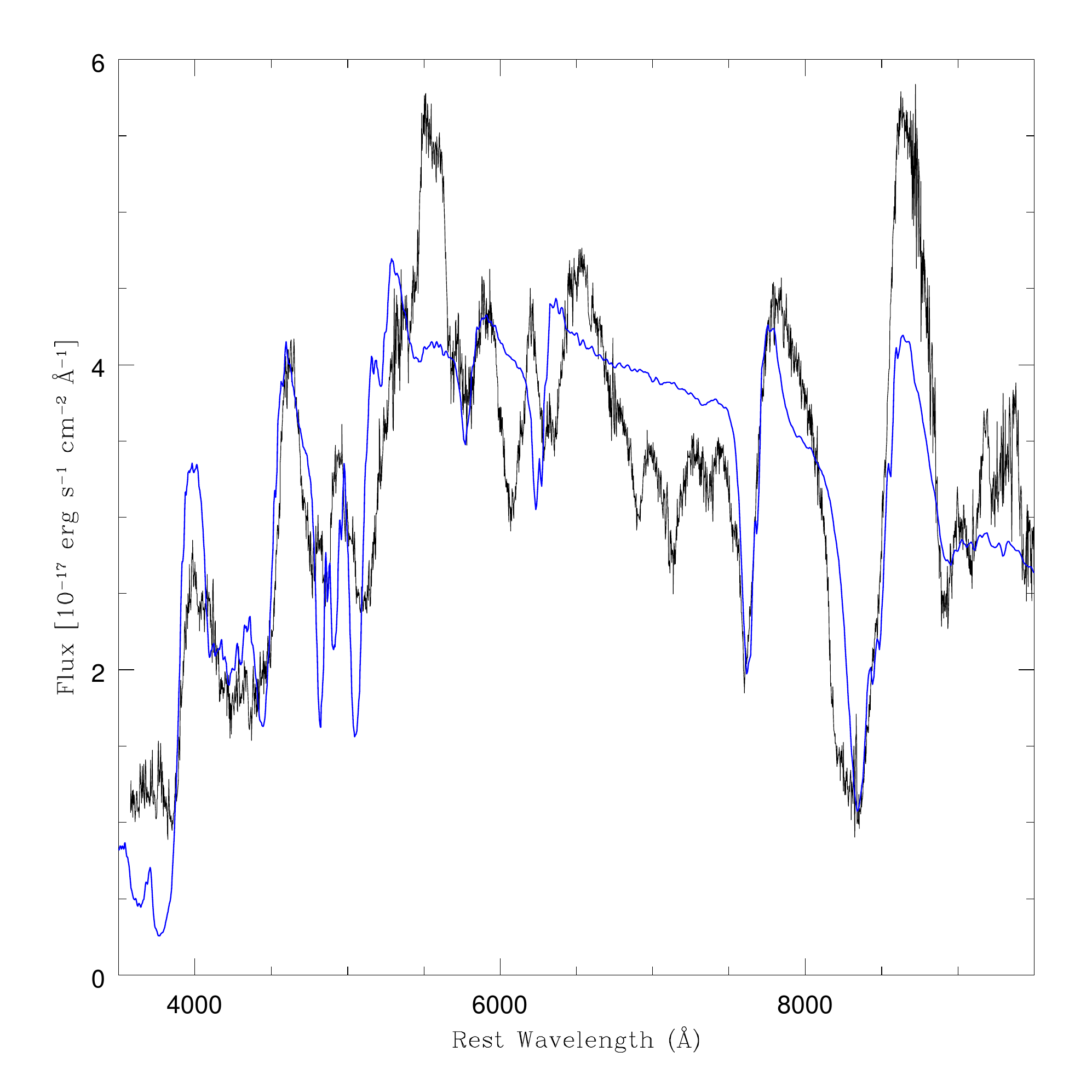}
\caption{Spectrum (black)  taken on 2012 Feb. 20 (72 rest-frame days after the explosion), 
flux-calibrated and  corrected as in Fig. \ref{fig:specsequence}.   The model shown as a blue curve  was obtained by applying a velocity cut to the density  profile at 15,000\,\kms.}
\label{fig:20Feb}
\end{figure*}

The spectrum at 72 rest-frame days past explosion is similar to the previous one, and has deep lines, especially Ca~II and O~I, but also Fe~II, owing presumably to the deep location of the photosphere. The best match we could find has $\log L = 42.66$\,[erg\,s$^{-1}$] and $\vph = 6000$\,\kms. The luminosity has not decreased very much in almost one month.  In general, the match is acceptable (Fig.\,\ref{fig:20Feb}), but we can clearly see the emergence of nebular emission, which  becomes stronger at later epochs. In particular, a strong emission line near 5500\,\AA\ is likely to be the [O~I] $\lambda$5577 line (a hint of it may be present already at day 49). The   near-infrared Ca~II triplet has a much stronger emission component, not in equilibrium with the absorption component, suggesting high-density emission.   

The spectral signal in the region $\sim 6000$--6500\,\AA\ is not well reproduced.  A  strong absorption feature, where the Si~II $\lambda\lambda6347$, 6371 doublet is expected, cannot be modelled satisfactorily despite using an increased Si abundance of 20\%  near the photosphere (which is obtained at the expense of carbon).  An emission feature is seen at the same location, which could be incipient nebular  [O~I] $\lambda\lambda6300$, 6364, although it appears to be shifted to the blue by at least 100\,\AA\  (and not asymmetric as would be the case if self-absorption were significant). It is also improbable that the entire absorption, on either side of the possible [O~I] $\lambda\lambda6300$, 6364 emission, could be Si~II.  The discrepancy may be due to the inability of our Monte Carlo code to  deal with this transitional spectrum in a  satisfactory way,  as some of the gas is already in the nebular regime.

\subsubsection{Spectrum at phase 157 d}
\label{sec:photspecmodel157d}

\begin{figure*} 
\includegraphics[width=0.8\textwidth]{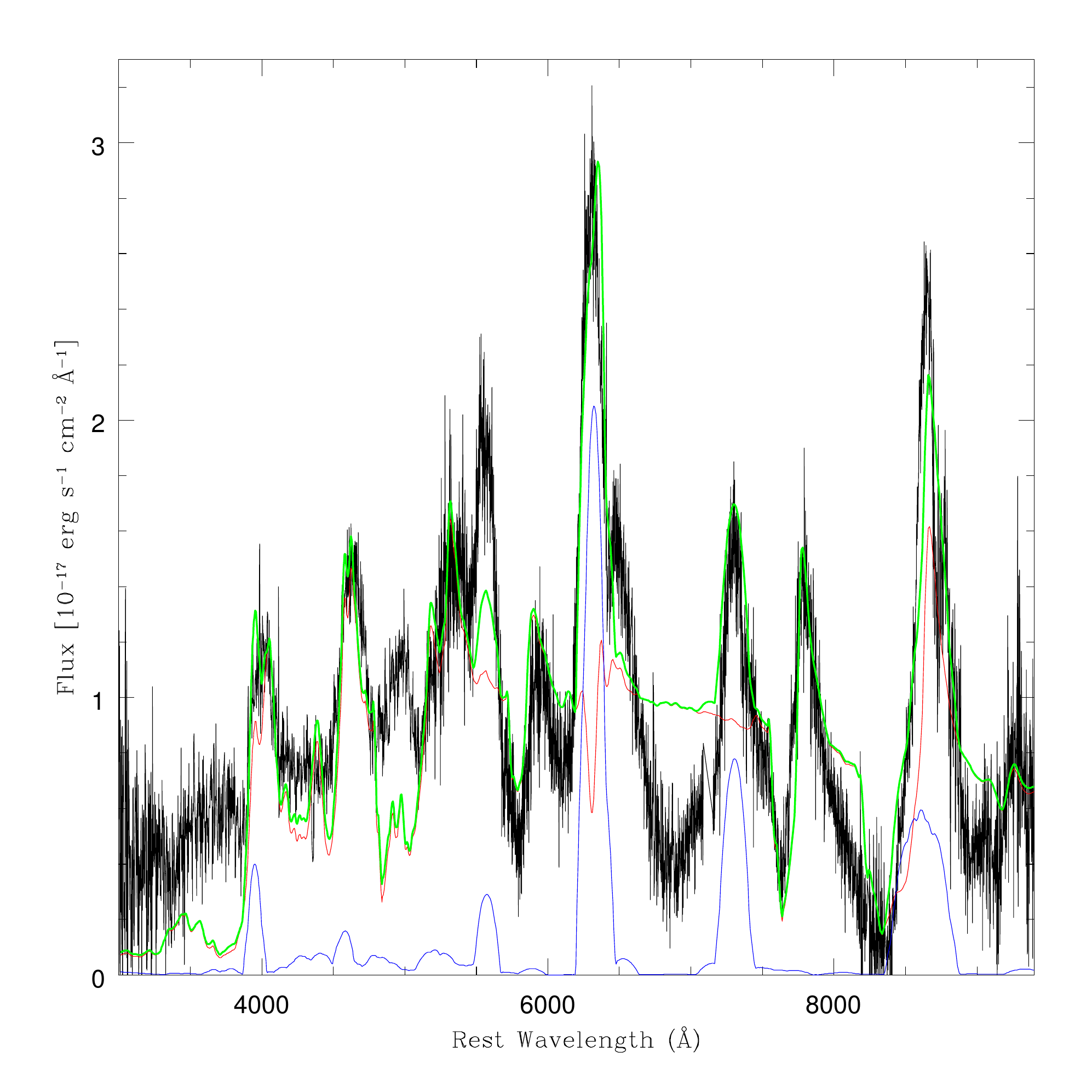}
\caption{Spectrum (black)  taken on 2012 May 22 (157 rest-frame days  after the explosion), 
 flux-calibrated and  corrected as in Fig. \ref{fig:specsequence}.   The model shown as a green curve is the sum of a photospheric (red curve) and nebular (blue curve) model (see text).}
\label{fig:22May}
\end{figure*}

Nebular emission clearly affects the following spectrum in the series (157 rest-frame  days after explosion), although many P-Cygni profiles can still be recognised (Fig. \ref{fig:22May}).  As our Monte Carlo code cannot deal with this hybrid regime, we resort to the strategy adopted for the advanced-epoch spectra of SN 1997ef by \citet{mazzali2000b}. We combined a synthetic spectrum computed under the photospheric approximation with a  spectrum computed with our nebular-phase code. 

The photospheric spectrum has $\log L = 42.05$\,[erg\,s$^{-1}$] and $\vph = 1250$\,\kms. This is a very low velocity and we cannot expect our code to work well with such a large mass (more than 7 \Msun) above the photosphere. The composition near the photosphere is similar to that of day 72. All of the carbon has now been replaced by silicon and oxygen, though we do not see the need for an increased iron abundance. 
Many of the observed features are reasonably well reproduced as P-Cygni lines. These include the Fe~II + Ti~II trough at 4500--5000\,\AA, the Na~I~D line, O~I $\lambda$7774, and the  near-infrared Ca~II triplet. The Si~II line is seen in the model, but it is swamped by [O~I] emission. 

The additional nebular emission is powered by  $\sim$ 0.06\,\Msun\ of \Nifs, with a limiting outer velocity of 5000\,\kms. This is a very simplistic approximation, but it serves the purpose of demonstrating how the spectrum can be formed. Some 6\,\Msun\ of oxygen would contribute to the emission, with smaller masses of Si, C, Ca, and so on. The main emission features are [O~I] $\lambda\lambda$6300, 6364, the  near-infrared Ca~II triplet, [Ca~II] $\lambda$7231, and [O~I] $\lambda$5577.

In Fig. \ref{fig:22May} it is shown that the sum of the  synthetic photospheric and nebular spectra  (green curve) matches the data  surprisingly well, although the approach has obviously no claim to perfect  consistency. The low optical depth region near 6800\,\AA, where  no flux is detected,  is not properly treated by our Monte Carlo code, which uses a lower boundary for the emission of photons.  The nebular spectrum contributes an additional $\log L = 41.15$\,[erg\,s$^{-1}$] to the total luminosity at that epoch, which is then $\log L = 42.10$\,[erg\,s$^{-1}$].

\subsubsection{Spectrum at phase 432 d}
\label{sec:nebspecmodel}

After subtraction of the host-galaxy spectral distribution from the VLT spectrum taken on 2013 March 11-15, the residual spectrum is very noisy and is affected by several weak and probably
spurious features (Fig. \ref{fig:vltnebspec}). However, the [O~I] $\lambda\lambda$6300, 6364 blend is
clearly visible, as is the [Ca~II] $\lambda$7231 emission. Among
several weak features, semiforbidden Mg~I] $\lambda$4571 appears to be present, and there are peaks that could match Na~I~D and [C~I] $\lambda$8727, the strongest expected optical  line of carbon. An important part of nebular modelling consists in matching Fe emission in a way that balances the inferred mass of \Nifs\ given the epoch. This is  not easy in this case as the spectrum is noisy.   Emission near 5200\,\AA\ matches the  [Fe~II] nebular line that is typical in SNe~Ib/c. Normally this emission is due to various lines, principally at 5159, 5262\,\AA, but in PTF11rka it appears narrower than expected given that the feature is typically a blend. The lack of well-identified Fe emission at this epoch makes the determination of the \Nifs\ mass through the combination of radioactive powering and decay quite uncertain. 

The [O~I] line is fairly narrow.  It can be matched with a boundary velocity for the nebula of 4000\,\kms, which is consistent with normal (i.e., not high $E_k$) SNe~Ic. As most cooling seems to take place in the [O~I] $\lambda\lambda$6300, 6364 line, we  started by matching that emission, and then iteratively added other elements to match other, weaker emission lines.  The main uncertainty in determining the ejected mass is the content of silicon and sulphur. These are the two most abundantly produced intermediate-mass elements. They can be ejected from the inner stellar core, but in the nebular phase they radiate mostly in the near-infrared, such that only if information in this wavelength band is available can one reliably quantify their production \citep[\eg,][]{mazzali2010,mazzali2015,mazzali2019}.   As near-infrared information is not available for PTF11rka we test two options: a high-mass solution, where as much as 2\,\Msun\ is Si and S in a ratio of 3:1, and a low-mass solution, where Si and S combined account for only 0.4\,\Msun, with the same ratio. A larger amount of intermediate-mass elements requires a greater \Nifs\ mass to heat the ejecta, keeping the oxygen mass roughly constant. 

Using an outer-boundary velocity of 4000\,\kms\  we find that a \Nifs\ mass of $\sim 0.45$\,\Msun\ yields good fits to the spectrum. The oxygen mass is $\sim 4$\,\Msun, the carbon mass is $\sim 1$\,\Msun, and the mass inside the outer boundary is $\sim 8$\,\Msun\ for the high-mass solution and $\sim 6$\,\Msun\ for the low-mass solution. A fairly large calcium mass was required to match the [Ca~II] $\lambda$7231 line (0.15\,\Msun). Small amounts of Mg (0.02\,\Msun) and Na (0.001\,\Msun) were also used. The Na~I~D line may be visible, but it seems very narrow in the data, so the Na mass carries great uncertainty. On the other hand, even large errors in the estimate of Ca and Na do not overly affect the overall energy balance, as these elements have a very small mass compared to oxygen. Therefore, our estimate of the total mass is mostly affected by the uncertain Si and S emission. 

The synthetic spectrum is shown superposed to the observed one in Fig.
\ref{fig:vltnebspec} (dark blue curve). The three strongest emission lines ([O~I], [Ca~II], and Mg~I]) are reproduced satisfactorily, but the [Fe~II] emission, although similar in strength to an observed spectral feature, is significantly broader,  and largely affected by noise.
Depending on the model we choose, the amount of luminosity needed to reproduce the spectrum is (2--4) $\times 10^{40}$\,erg\,s$^{-1}$. This can be used as an estimate for the bolometric luminosity at this late epoch.

We also tried to reproduce the spectrum using the density structure defined in Section \ref{sec:photspecmodel49d}.   The poor signal-to-noise ratio does not allow us to determine the inner density structure with a high level of confidence, but it confirms that a reasonable match can be obtained for a \Nifs\ mass of $\sim 0.35$\,\Msun.

Despite the large uncertainties, two main conclusions can be drawn from the nebular modelling.  First, the ejected mass is fairly large, but not comparable to that of events such as SN~2007bi. We can compare the ejecta of PTF11rka to those of SN 1998bw and approximate a total ejected mass of $\sim 8$\,\Msun.   Second, the role of \Nifs\ in supporting the SN luminosity cannot be neglected. We estimate  that $\sim 0.4$\,\Msun\ of \Nifs\ were ejected by the SN, comparable to SN  1998bw. This may confirm a tendency for the \Nifs\ mass to grow with the ejected mass.

\subsection{Light curve}
\label{sec:lcmodel}

In the last step of our modelling procedure, we compute a synthetic bolometric light curve using the density structure derived from the spectral modelling. We regard the luminosities that went into the spectral fits as the actual  bolometric values, as quantities determined from the photometry (Section \ref{sec:obsphotom}) are highly uncertain owing to poor photometric coverage as well as unknown contribution of  flux outside the observed bands. 

We used our Monte Carlo light-curve code \citep{cappellaro1997}. The code follows the emission and diffusion of gamma rays and positrons from \Nifs\ and \Cofs\ decay, based on a radial distribution of \Nifs\ and a one-dimensional density structure. Gamma-ray deposition occurs based on an opacity $\kappa_{\gamma} = 0.027$\,cm$^2$\,g$^{-1}$, while positron deposition is computed based on an opacity approximation with $\kappa_{e^+} = 7$\,cm$^2$\,g$^{-1}$. Upon deposition of gamma rays and positrons, packets of optical energy are assumed to be emitted, which then diffuse and are subject to a composition-dependent line opacity, as set out by \citet{mazzali2001b}. This assumes that line opacity is the dominant form of opacity in H-free SNe \citep{pauldrach1996}.

\begin{figure*} 
\includegraphics[width=0.8\textwidth]{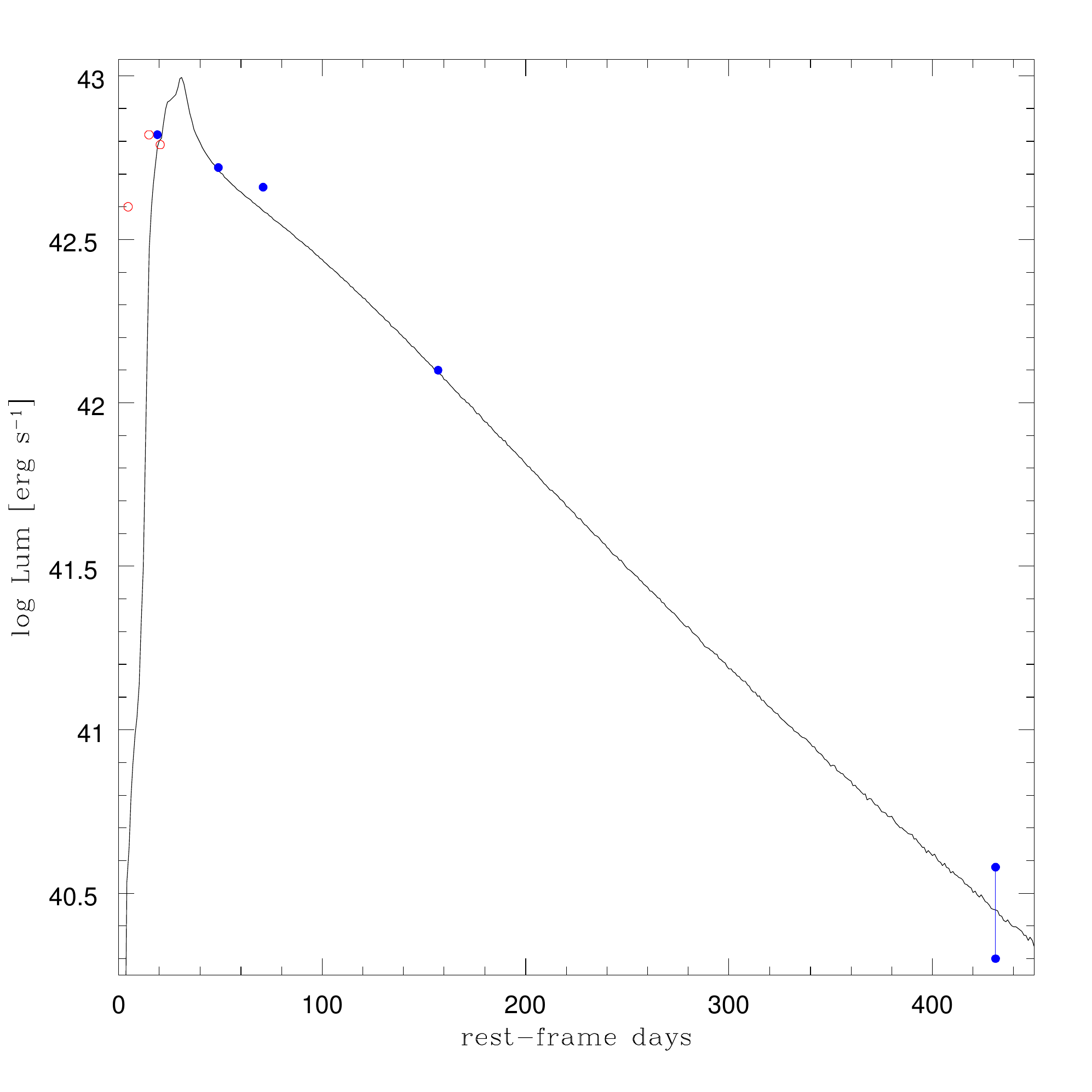}
\caption{Light-curve model (black curve).  The blue filled circles represent the input luminosities of the spectral models.  The vertical bar at 432 days represents the range of luminosities covered by the two adopted mass solutions for the nebular phase (Section \ref{sec:nebspecmodel}).  The open red circles represent the pre-maximum bolometric light-curve points, obtained by multiplying  the $R_{\rm PTF}$ fluxes by a constant amount equivalent to the ratio between the model luminosity at day 20d and the simultaneous $R_{\rm PTF}$ flux.}
\label{fig:lcmodel}
\end{figure*}

We can obtain a reasonable fit to the light curve (which only has five points), but we need to use a larger \Nifs\ mass  ($\sim 0.5$\,\Msun) than what was used for the nebular spectrum ($\sim 0.35$\,\Msun).  This is due both to the very poor quality of the nebular spectrum on day 432, where the Fe lines are not well defined, and to the different assumptions on the geometrical distribution of \Nifs\  made here and in Section \ref{sec:nebspecmodel}. Our synthetic light curve is shown in Fig.\ref{fig:lcmodel}.   The high \Nifs\ mass is necessary to match the luminosity near maximum brightness, which is estimated to occur $\sim 30$ days after the explosion at a luminosity of $\sim 10^{43}$\,erg\,s$^{-1}$.   A significant fraction of this mass ($\sim 0.15$\,\Msun) is located at velocities larger than 4000\,\kms, and thus does not contribute to the light curve at nebular epochs, hence the (modest) \Nifs\ mass discrepancy.

Our model light curve fails to reproduce the few points  prior to the first spectral observation. These (shown as open red circles in Fig. \ref{fig:lcmodel})  were obtained by applying a bolometric correction to the $R_{\rm PTF}$ measurements equivalent to the difference between the spectral model luminosity at 20\,days and the simultaneous  luminosity in the $R_{\rm PTF}$ band.   If these points are real, they would require an unreasonably large amount of \Nifs\ at high velocity to guarantee an early rise of the light curve. This, however, is not corroborated by the early-time spectral models.  Alternatively, the early points could reflect the conversion of some SN \KE\ into radiative energy. This conversion would be mediated by the impact of the outer ejecta with CSM, as was already inferred from the narrow absorptions seen in the early-time spectra.

As mentioned above, the amount of \KE\ removed from the explosion model in order to achieve narrow lines was $\sim 5 \times 10^{50}$\,erg. The amount of luminosity in the earliest phase of the light curve may be estimated as $\sim 5 \times 10^{42}$\,erg\,s$^{-1}$ for $\sim 20$\,d, i.e. $\sim 10^{49}$\,erg. This amount of radiative energy is only $\sim 2$\% of the \KE\ spent in the interaction, a plausible conversion efficiency. 

Note that the model light curve favours maximum light at $\sim 30$\,days after explosion, and coincident with $R$-band maximum (Figs. \ref{fig:mwllcs} and \ref{fig:lcmodel}).
This underlines the fact that the pseudobolometric light curve (Fig. \ref{fig:3bollc}) is  unreliable, as it possibly seriously underestimates contributions from unobserved bands.


\section{Summary and Conclusions}
\label{sec:discussion}

The somewhat unusual properties of PTF11rka can be explained if a
combination of events took place that make it rather unique, but at the same time
offer a useful link to other, more famous SNe.  First,  the spectra
and most of the light curve can be explained to a large extent in the
traditional picture where the SN ejects a massive stellar core and the luminosity is supported by the
radioactive decay of \Nifs. A fairly large \Nifs\ mass is required, $\sim
0.4$--0.5\,\Msun, comparable to the ejection in GRB-SNe,   in particular the prototype of this class,
SN~1998bw \citep{nakamura2001,woosleybloom2006,ashall2019}.   Also, the estimated ejecta mass,  \Mej\ $\approx 8 \pm 2$\,\Msun, is on the high side of the distribution of stripped-envelope core-collapse SNe \citep{ashall2019}.  From \Mej\  we infer a CO stellar core mass in the range $M \approx 8$--13\,\Msun\ --  depending on whether the remnant is a low-mass neutron star or  a black hole  -- resulting from the evolution of a  progenitor star  of  main-sequence mass 25--40\,\Msun\  \citep{nomotohashimoto1988,heger2003}.    On the other hand,  the kinetic energy we derived  (\KE\ $\approx 4 \times 10^{51}$\,erg),  although high,  is not extreme and more similar to that of energetic stripped-envelope SNe not accompanied by GRBs \citep{mazzali2017,ashall2019}.  This, together with the high \Mej\, results in a rather low  \KE/\Mej\ ratio ($\sim 0.4$).

However, the early-time spectra of PTF11rka resemble those of the SLSN~2007bi.
In recent work, \citet{moriya2019b} and \citet{mazzali2019} showed that SN~2007bi is consistent with the explosion of a $\sim 40$\,\Msun\ CO core of a massive star (of initial mass probably 60--80\,\Msun). The explosion was not particularly energetic, given the large mass (\KE\  $\approx  4 \times 10^{52}$\,erg, \KE/\Mej\ $\approx 1$), which makes the light curve broad. A best match to the spectra of SN~2007bi (and SN~1999as) was obtained if it was assumed that the outer, highest velocity layers of even this low-\KE\ explosion were ``cut'', which physically is likely to mean that they were slowed down in an impact with CSM. A similar solution holds for PTF11rka. The density distribution of the  ejecta had to be cut at a velocity of 15,000\,\kms, which corresponds to $\sim 0.1$\,\Msun, with the loss of $\sim 5 \times 10^{50}$\,erg of kinetic energy.  When the suppression is applied a much better reproduction of the observed early-time spectra is obtained.   The \KE\ that was lost upon interaction may have partly been converted to radiative luminosity, which could have led to a rapid rise of the light curve, before radioactive decay started playing a major role. 

Comparison of the peak luminosity of PTF11rka  with those of H-poor SNe   \citep[see][]{decia2018} shows that the former is less luminous than SLSNe by about  an order of magnitude, while it sits squarely in the range of Type Ib/c SNe, and at the low-luminosity end of broad-lined SNe~Ic, when due account is taken for the fact that our  pseudobolometric luminosity estimate for PTF11rka is based only on  optical data (with no correction for possible ultraviolet or infrared contributions).   In particular, our  estimated \Nifs\ mass of PTF11rka is very close to that of GRB-associated SN 1998bw, and  both its broad light-curve shape and peak luminosity are similar to those of the ``spectroscopically normal'' Type Ic SN 2011bm \citep{valenti2012},  as are the physical parameters (\KE, \Mej,  \Nifs\ mass, and progenitor mass of PTF11rka resemble the lower boundaries of the parameter ranges estimated for SN 2011bm).
From the spectral point of view, while PTF11rka is spectroscopically similar to SLSN~2007bi at early epochs, its late-time spectra (days 157 and 432 after explosion)  are reminiscent of those of H-poor normal SNe, broad-lined SNe, and SLSNe \citep{pastorello2010,jerkstrand2017}.

These facts make PTF11rka a gap-bridging  object between normal and energetic  stripped-envelope SNe (including GRB-SNe)  and H-poor SLSNe.
Furthermore,  the analogy of PTF11rka with SLSN~2007bi demonstrates that ignoring both \Nifs\ and CSM
interaction in H-poor SLSNe may be an oversimplification. Efforts should be made to
disentangle these various components. Availability of early and late-time data,
accompanied by a careful analysis, are powerful steps to improve our
understanding of SLSNe, as the example of the non-SLSN PTF11rka shows.

\section*{Acknowledgements}

We are grateful to Y. Cao,  J. Bloom, and J. M. Silverman for their assistance with data acquisition and reduction.   E.P. thanks the Weizmann Institute for Science (Rehovot, Israel), the National Astronomical Observatory of Japan, the Astrophysics Research Institute of Liverpool John Moores University,  
and  the Munich Institute for Astrophysics and Particle Physics (MIAPP) for hospitality and support, and acknowledges fruitful conversations and exchanges with the participants of  MIAPP programs  ``The Physics of Supernovae'' (2016) and ``Superluminous Supernovae in the Next Decade'' (2017).   
T.J.M. is supported by the Grants-in-Aid for Scientific Research of the Japan Society for the Promotion of Science (JP17H02864, JP18K13585).    
A.G-Y.'s research is supported by the EU via ERC grant 725161, the ISF GW Excellence Center, an IMOS space infrastructure grant, and BSF/Transformative and GIF grants, as well as by the Benoziyo Endowment Fund for the Advancement of Science, the Deloro Institute for Advanced Research in Space and Optics, The Veronika A. Rabl Physics Discretionary Fund, Paul and Tina Gardner, Yeda-Sela, and the WIS-CIT joint research grant;  A.G.-Y. is also the recipient of the Helen and Martin Kimmel Award for Innovative Investigation.
I.A. is a CIFAR Azrieli Global Scholar in the Gravity and the Extreme Universe Program and acknowledges support from that program, from the European Research Council (ERC) under the European UnionÕs Horizon 2020 research and innovation program (grant agreement number 852097), from the Israel Science Foundation (grant numbers 2108/18 and 2752/19), from the United States - Israel Binational Science Foundation (BSF), and from the Israeli Council for Higher Education Alon Fellowship.
A.V.F. is grateful for financial assistance from U.S. National Science Foundation grant AST-1211916, the  TABASGO Foundation, the Christopher R. Redlich Fund, and the Miller Institute for Basic Research in Science (U.C. Berkeley).
This research was supported  by the Italian Ministry for  Research (PRIN MIUR 2010/2011), INAF (PRIN INAF 2011 and 2014),  by Scuola Normale Superiore, and by the NAOJ Research Coordination Committee, NINS, grant numbers 19FS-0506 and 19FS-0507.

IRAF is distributed by the National Optical Astronomy Observatory, which is operated by the Association of Universities for Research in Astronomy (AURA), Inc., under a cooperative agreement with the U.S. NSF.   This research has made use of the NASA/IPAC Extragalactic Database (NED) which is operated by the Jet Propulsion Laboratory, California Institute of Technology, under contract with the National Aeronautics and Space Administration (NASA).
Some of the data presented herein were obtained at the W. M. Keck Observatory, which is operated as a scientific partnership among the California Institute of Technology, the University of California, and NASA; the observatory was made possible by the generous financial support of the W. M. Keck Foundation.

This paper is dedicated to the memory of our friend and colleague Adi Pauldrach.

\section*{Data availability}

The photometric and spectroscopic data presented in this article are publicly available via the Weizmann Interactive Supernova Data Repository \citep{yarongalyam2012},  at  https://wiserep.weizmann.ac.il.




\bibliographystyle{mnras}
\bibliography{example} 




\begin{table}
\centering 
\caption{Log of PTF11rka photometry.} 
\label{tab:logphotomobs}
\begin{tabular}{c c c c c } 
\hline
\hline
Date & MJD$^a$ & Phase$^b$                 &  Filter &  mag$^c$ \\
(UT)   &          &  (days) &           &           \\ 
\hline
 2011 Nov 16.01  & 55881.51    &   -17.20    &  $R_{PTF}$ &   $>$ 20.77  \\
 2011 Nov  16.02  &55881.52     &  -17.20    &  $R_{PTF}$ &   $>$ 20.82  \\
 2011 Nov  22.99  &55888.49     &  -10.72    &  $R_{PTF}$ &   $>$ 20.60  \\
 2011 Nov  22.52  & 55888.52    &  -10.69    &  $R_{PTF}$ &   $>$ 20.71  \\
 2011 Nov  26.00  & 55891.50    &   -7.92  &  $R_{PTF}$ &   $>$ 20.26  \\
 2011 Nov  26.04  & 55891.54    &  -7.88   &  $R_{PTF}$ &   $>$ 21.04  \\
 2011 Nov 30.49   &  55895.48  &   -4.20   &   $R_{PTF}$ &   $>$ 21.23  \\  
 2011 Nov 30.53   &  55895.53  &   -4.16   &   $R_{PTF}$ &   $>$ 21.32  \\  
 2011 Dec  4.46   &  55899.46  &  -0.51   &   $R_{PTF}$ &   $>$ 20.58  \\ 
 2011  Dec 4.50   & 55899.50   &   -0.46    &   $R_{PTF}$ &   $>$ 20.57  \\
 2011 Dec  7.45  &  55902.45  &    2.28  & $R_{PTF}$ &  20.05 $\pm$ 0.31  \\ 
 2011 Dec  7.50  &  55902.50  &    2.32  & $R_{PTF}$ &  19.98 $\pm$ 0.14  \\ 
 2011 Dec 10.45  &  55905.45  &    5.07  & $R_{PTF}$ &  19.53 $\pm$ 0.28  \\ 
 2011 Dec 10.49  &  55905.49  &    5.11  & $R_{PTF}$ &  19.34 $\pm$ 0.43  \\ 
 2011 Dec 21.41  &  55916.41  &   15.28  & $R_{PTF}$ &  18.95 $\pm$ 0.06  \\ 
 2011 Dec 21.46  &  55916.46  &   15.32  & $R_{PTF}$ &  18.95 $\pm$ 0.05  \\ 
 2011 Dec 27.42  &  55922.42  &   20.87  & $R_{PTF}$ &  19.05 $\pm$ 0.05  \\ 
 2011 Dec 27.47  &  55922.46  &   20.91  & $r$ &  18.97 $\pm$ 0.02  \\ 
 2011 Dec 27.47  &  55922.46  &   20.91  & $R_{PTF}$ &  18.89 $\pm$ 0.04  \\ 
 2011 Dec 27.50  &  55922.50  &   20.95  & $i$ &  19.00 $\pm$ 0.03  \\ 
 2011 Dec 27.50  &  55922.50  &   20.95  & $r$ &  18.98 $\pm$ 0.02  \\ 
 2011 Dec 27.51  &  55922.51  &   20.95  & $g$ &  19.42 $\pm$ 0.02  \\ 
 2011 Dec 28.39  &  55923.39  &   21.77  & $i$ &  18.83 $\pm$ 0.05  \\ 
 2011 Dec 28.39  &  55923.39  &   21.77  & $r$ &  19.01 $\pm$ 0.04  \\ 
 2011 Dec 28.40  &  55923.40  &   21.78  & $g$ &  19.54 $\pm$ 0.06  \\ 
 2011 Dec 30.39  &  55925.39  &   23.64  & $R_{PTF}$ &  18.92 $\pm$ 0.04  \\ 
 2011 Dec 30.44  &  55925.44  &   23.68  & $R_{PTF}$ &  18.84 $\pm$ 0.04  \\ 
 2011 Dec 31.39  &  55926.39  &   24.56  & $i$ &  18.95 $\pm$ 0.08  \\ 
 2011 Dec 31.39  &  55926.39  &   24.56  & $r$ &  18.98 $\pm$ 0.09  \\ 
 2011 Dec 31.39  &  55926.39  &   24.57  & $g$ &  19.61 $\pm$ 0.14  \\ 
 2012 Jan  2.38  &  55928.38  &   26.41  & $R_{PTF}$ &  18.85 $\pm$ 0.06  \\ 
 2012 Jan  2.43  &  55928.42  &   26.45  & $R_{PTF}$ &  18.92 $\pm$ 0.05  \\ 
 2012 Jan  3.38  &  55929.38  &   27.34  & $i$ &  18.91 $\pm$ 0.04  \\ 
 2012 Jan  3.38  &  55929.38  &   27.34  & $r$ &  18.91 $\pm$ 0.04  \\ 
 2012 Jan  3.38  &  55929.38  &   27.34  & $g$ &  19.81 $\pm$ 0.09  \\ 
 2012 Jan  5.37  &  55931.38  &   29.20  & $R_{PTF}$ &  18.88 $\pm$ 0.26  \\ 
 2012 Jan  5.39  &  55931.39  &   29.22  & $R_{PTF}$ &  18.81 $\pm$ 0.11  \\ 
 2012 Jan 10.36  &  55936.36  &   33.84  & $R_{PTF}$ &  18.95 $\pm$ 0.11  \\ 
 2012 Jan 10.38  &  55936.38  &   33.86  & $R_{PTF}$ &  18.73 $\pm$ 0.14  \\ 
 2012 Jan 17.34  &  55943.34  &   40.34  & $i$ &  18.88 $\pm$ 0.05  \\ 
 2012 Jan 17.34  &  55943.34  &   40.34  & $r$ &  19.03 $\pm$ 0.04  \\ 
 2012 Jan 17.35  &  55943.35  &   40.35  & $g$ &  20.10 $\pm$ 0.08  \\ 
 2012 Jan 17.42  &  55943.42  &   40.41  & $R_{PTF}$ &  18.95 $\pm$ 0.05  \\ 
 2012 Jan 17.42  &  55943.42  &   40.42  & $R_{PTF}$ &  18.96 $\pm$ 0.05  \\ 
 2012 Jan 17.42  &  55943.42  &   40.42  & $R_{PTF}$ &  18.97 $\pm$ 0.05  \\ 
 2012 Jan 20.33  &  55946.33  &   43.12  & $i$ &  18.96 $\pm$ 0.06  \\ 
 2012 Jan 20.33  &  55946.33  &   43.12  & $R_{PTF}$ &  19.00 $\pm$ 0.08  \\ 
 2012 Jan 20.34  &  55946.34  &   43.13  & $r$ &  19.07 $\pm$ 0.06  \\ 
 2012 Jan 20.34  &  55946.34  &   43.13  & $g$ &  20.09 $\pm$ 0.11  \\ 
 2012 Jan 25.41  &  55951.41  &   47.85  & $i$ &  18.95 $\pm$ 0.07  \\ 
 2012 Jan 25.41  &  55951.41  &   47.85  & $r$ &  19.22 $\pm$ 0.05  \\ 
 2012 Jan 25.41  &  55951.41  &   47.85  & $g$ &  20.23 $\pm$ 0.10  \\ 
 2012 Jan 25.43  &  55951.43  &   47.86  & $R_{PTF}$ &  18.96 $\pm$ 0.06  \\ 
 2012 Jan 28.31  &  55954.32  &   50.56  & $g$ &  20.30 $\pm$ 0.20  \\ 
 2012 Jan 28.38  &  55954.38  &   50.61  & $R_{PTF}$ &  19.18 $\pm$ 0.10  \\ 
 2012 Jan 28.42  &  55954.42  &   50.65  & $R_{PTF}$ &  19.19 $\pm$ 0.09  \\ 
 2012 Jan 31.37  &  55957.37  &   53.40  & $g$ &  20.45 $\pm$ 0.06  \\ 
 2012 Feb  1.30  &  55958.30  &   54.26  & $r$ &  19.36 $\pm$ 0.04  \\ 
 2012 Feb  1.30  &  55958.30  &   54.26  & $R_{PTF}$ &  19.12 $\pm$ 0.08  \\ 
 2012 Feb  1.30  &  55958.30  &   54.26  & $i$ &  19.09 $\pm$ 0.04  \\ 
\hline
\end{tabular}
\end{table}
\begin{table}
\begin{tabular}{c c c c c } 
\multicolumn{5}{l}{{\bf Table 1} (Continued).} \\
\hline
\hline
Date & MJD$^a$ & Phase$^b$                 &  Filter &  mag$^c$ \\
(UT)   &          &  (days) &           &           \\ 
\hline 
 2012 Feb  1.34  &  55958.34  &   54.30  & $R_{PTF}$ &  19.15 $\pm$ 0.06  \\ 
 2012 Feb  3.29  &  55960.29  &   56.12  & $g$ &  20.50 $\pm$ 0.37  \\ 
 2012 Feb  4.36  &  55961.36  &   57.11  & $R_{PTF}$ &  19.26 $\pm$ 0.10  \\ 
 2012 Feb  4.40  &  55961.40  &   57.15  & $R_{PTF}$ &  19.32 $\pm$ 0.08  \\ 
 2012 Feb  5.29  &  55962.28  &   57.97  & $i$ &  19.20 $\pm$ 0.06  \\ 
 2012 Feb  5.29  &  55962.29  &   57.97  & $r$ &  19.38 $\pm$ 0.07  \\ 
 2012 Feb  5.29  &  55962.29  &   57.98  & $g$ &  20.77 $\pm$ 0.25  \\ 
 2012 Feb  6.36  &  55963.36  &   58.97  & $i$ &  19.23 $\pm$ 0.12  \\ 
 2012 Feb  6.36  &  55963.36  &   58.98  & $r$ &  19.29 $\pm$ 0.13  \\ 
 2012 Feb  6.37  &  55963.37  &   58.98  & $g$ &  20.34 $\pm$ 0.18  \\ 
 2012 Feb 17.26  &  55974.25  &   69.11  & $r$ &  19.82 $\pm$ 0.16  \\ 
 2012 Feb 17.25  &  55974.25  &   69.11  & $i$ &  19.33 $\pm$ 0.13  \\ 
 2012 Feb 18.25  &  55975.25  &   70.04  & $r$ &  19.59 $\pm$ 0.05  \\ 
 2012 Feb 18.25  &  55975.25  &   70.04  & $g$ &  20.75 $\pm$ 0.10  \\ 
 2012 Feb 18.27  &  55975.26  &   70.05  & $R_{PTF}$ &  19.73 $\pm$ 0.11  \\ 
 2012 Feb 18.31  &  55975.31  &   70.09  & $R_{PTF}$ &  19.59 $\pm$ 0.08  \\ 
 2012 Feb 20.44  &  55977.44  &   72.08  & $i$ &  19.37 $\pm$ 0.06  \\ 
 2012 Feb 20.44  &  55977.44  &   72.08  & $i$ &  19.46 $\pm$ 0.06  \\ 
 2012 Feb 21.38  &  55978.38  &   72.95  & $R_{PTF}$ &  19.57 $\pm$ 0.08  \\ 
 2012 Feb 24.29  &  55981.29  &   75.66  & $R_{PTF}$ &  19.61 $\pm$ 0.08  \\ 
 2012 Feb 24.33  &  55981.33  &   75.70  & $R_{PTF}$ &  19.67 $\pm$ 0.08  \\ 
 2012 Feb 26.23  &  55983.23  &   77.47  & $g$ &  20.91 $\pm$ 0.12  \\ 
 2012 Feb 29.48  &  55986.48  &   80.49  & $R_{PTF}$ &  19.35 $\pm$ 0.26  \\ 
 2012 Mar  2.42  &  55988.42  &   82.30  & $r$ &  19.84 $\pm$ 0.06  \\ 
 2012 Mar  2.42  &  55988.42  &   82.30  & $i$ &  19.67 $\pm$ 0.11  \\ 
 2012 Mar  2.42  &  55988.42  &   82.30  & $g$ &  20.82 $\pm$ 0.13  \\ 
 2012 Mar  4.26  &  55990.26  &   84.01  & $R_{PTF}$ &  19.81 $\pm$ 0.19  \\ 
 2012 Mar  4.30  &  55990.30  &   84.05  & $R_{PTF}$ &  19.96 $\pm$ 0.18  \\ 
 2012 Mar  6.36  &  55992.36  &   85.97  & $i$ &  19.84 $\pm$ 0.10  \\ 
 2012 Mar  6.37  &  55992.37  &   85.97  & $r$ &  19.97 $\pm$ 0.11  \\ 
 2012 Mar  6.37  &  55992.37  &   85.97  & $g$ &  20.70 $\pm$ 0.23  \\ 
 2012 Mar 13.23  &  55999.23  &   92.36  & $R_{PTF}$ &  19.82 $\pm$ 0.09  \\ 
 2012 Mar 13.28  &  55999.28  &   92.40  & $R_{PTF}$ &  19.89 $\pm$ 0.08  \\ 
 2012 Mar 13.34  &  55999.34  &   92.46  & $R_{PTF}$ &  20.15 $\pm$ 0.15  \\ 
 2012 Mar 13.38  &  55999.38  &   92.50  & $R_{PTF}$ &  19.79 $\pm$ 0.11  \\ 
 2012 Mar 13.46  &  55999.46  &   92.57  & $R_{PTF}$ &  19.64 $\pm$ 0.10  \\ 
 2012 Mar 13.47  &  55999.47  &   92.58  & $R_{PTF}$ &  20.03 $\pm$ 0.14  \\ 
 2012 Mar 14.33  &  56000.33  &   93.38  & $R_{PTF}$ &  20.05 $\pm$ 0.10  \\ 
 2012 Mar 14.37  &  56000.38  &   93.42  & $R_{PTF}$ &  19.85 $\pm$ 0.10  \\ 
 2012 Mar 14.44  &  56000.44  &   93.49  & $R_{PTF}$ &  20.07 $\pm$ 0.12  \\ 
 2012 Mar 14.48  &  56000.48  &   93.52  & $R_{PTF}$ &  19.76 $\pm$ 0.10  \\ 
 2012 Mar 15.23  &  56001.23  &   94.22  & $R_{PTF}$ &  20.25 $\pm$ 0.13  \\ 
 2012 Mar 15.27  &  56001.27  &   94.26  & $R_{PTF}$ &  19.94 $\pm$ 0.09  \\ 
 2012 Mar 15.34  &  56001.34  &   94.32  & $R_{PTF}$ &  20.11 $\pm$ 0.11  \\ 
 2012 Mar 15.38  &  56001.38  &   94.36  & $R_{PTF}$ &  19.89 $\pm$ 0.09  \\ 
 2012 Mar 16.25  &  56002.24  &   95.16  & $R_{PTF}$ &  20.07 $\pm$ 0.09  \\ 
 2012 Mar 16.26  &  56002.26  &   95.18  & $R_{PTF}$ &  19.91 $\pm$ 0.08  \\ 
 2012 Mar 16.33  &  56002.32  &   95.24  & $R_{PTF}$ &  19.86 $\pm$ 0.07  \\ 
 2012 Mar 16.34  &  56002.34  &   95.25  & $R_{PTF}$ &  19.96 $\pm$ 0.08  \\ 
 2012 Mar 17.22  &  56003.22  &   96.07  & $R_{PTF}$ &  19.88 $\pm$ 0.22  \\ 
 2012 Mar 20.47  &  56006.46  &   99.09  & $R_{PTF}$ &  19.96 $\pm$ 0.10  \\ 
 2012 Mar 21.17  &  56007.16  &   99.74  & $R_{PTF}$ &  19.90 $\pm$ 0.15  \\ 
 2012 Mar 21.21  &  56007.21  &   99.78  & $R_{PTF}$ &  20.36 $\pm$ 0.16  \\ 
 2012 Mar 21.47  &  56007.46  &  100.02  & $R_{PTF}$ &  20.13 $\pm$ 0.12  \\ 
 2012 Mar 22.16  &  56008.16  &  100.67  & $R_{PTF}$ &  19.95 $\pm$ 0.12  \\ 
 2012 Mar 22.21  &  56008.21  &  100.71  & $R_{PTF}$ &  20.06 $\pm$ 0.10  \\ 
 2012 Mar 22.27  &  56008.27  &  100.77  & $R_{PTF}$ &  20.00 $\pm$ 0.09  \\ 
 2012 Mar 22.31  &  56008.32  &  100.82  & $R_{PTF}$ &  19.99 $\pm$ 0.08  \\ 
 2012 Mar 23.17  &  56009.17  &  101.61  & $R_{PTF}$ &  19.90 $\pm$ 0.10  \\ 
 2012 Mar 23.21  &  56009.21  &  101.64  & $R_{PTF}$ &  20.02 $\pm$ 0.09  \\ 
\hline
\end{tabular}
\end{table}
\begin{table}
\begin{tabular}{c c c c c } 
\multicolumn{5}{l}{{\bf Table 1} (Continued).} \\
\hline
\hline
Date & MJD$^a$ & Phase$^b$                 &  Filter &  mag$^c$ \\
(UT)   &          &  (days) &           &           \\ 
\hline 
 2012 Mar 23.47  &  56009.46  &  101.88  & $R_{PTF}$ &  20.04 $\pm$ 0.09  \\ 
 2012 Mar 23.51  &  56009.50  &  101.92  & $R_{PTF}$ &  20.04 $\pm$ 0.12  \\ 
 2012 Mar 28.19  &  56014.20  &  106.29  & $R_{PTF}$ &  19.96 $\pm$ 0.19  \\ 
 2012 Mar 28.27  &  56014.27  &  106.36  & $R_{PTF}$ &  19.96 $\pm$ 0.20  \\ 
 2012 Mar 29.19  &  56015.18  &  107.21  & $i$ &  20.20 $\pm$ 0.09  \\ 
 2012 Mar 29.19  &  56015.19  &  107.21  & $r$ &  20.26 $\pm$ 0.08  \\ 
 2012 Mar 29.19  &  56015.20  &  107.22  & $g$ &  20.97 $\pm$ 0.17  \\ 
 2012 Mar 31.17  &  56017.18  &  109.06  & $r$ &  20.22 $\pm$ 0.08  \\ 
 2012 Mar 31.17  &  56017.18  &  109.06  & $i$ &  20.13 $\pm$ 0.09  \\ 
 2012 Mar 31.18  &  56017.18  &  109.07  & $g$ &  21.40 $\pm$ 0.24  \\ 
 2012 Apr  3.17  &  56020.17  &  111.85  & $i$ &  20.13 $\pm$ 0.24  \\ 
 2012 Apr  3.17  &  56020.17  &  111.85  & $r$ &  20.11 $\pm$ 0.19  \\ 
 2012 Apr  4.17  &  56021.17  &  112.78  & $r$ &  20.10 $\pm$ 0.13  \\ 
 2012 Apr  4.17  &  56021.17  &  112.78  & $i$ &  19.98 $\pm$ 0.14  \\ 
 2012 Apr  8.17  &  56025.18  &  116.51  & $i$ &  20.10 $\pm$ 0.16  \\ 
 2012 Apr  8.18  &  56025.18  &  116.51  & $r$ &  20.36 $\pm$ 0.17  \\ 
 2012 Apr  9.16  &  56026.16  &  117.42  & $r$ &  20.23 $\pm$ 0.06  \\ 
 2012 Apr  9.16  &  56026.16  &  117.42  & $i$ &  20.22 $\pm$ 0.06  \\ 
 2012 Apr  9.16  &  56026.16  &  117.43  & $g$ &  21.39 $\pm$ 0.15  \\ 
 2012 Apr 16.19  &  56033.19  &  123.97  & $i$ &  20.34 $\pm$ 0.07  \\ 
 2012 Apr 16.19  &  56033.20  &  123.97  & $r$ &  20.32 $\pm$ 0.05  \\ 
 2012 Apr 16.20  &  56033.20  &  123.98  & $g$ &  21.46 $\pm$ 0.13  \\ 
 2012 Apr 19.45  &  56036.44  &  126.99  & $r$ &  20.50 $\pm$ 0.11  \\ 
 2012 Apr 19.44  &  56036.44  &  126.99  & $i$ &  20.33 $\pm$ 0.14  \\ 
 2012 Apr 19.46  &  56036.46  &  127.01  & $g$ &  21.11 $\pm$ 0.26  \\ 
 2012 Apr 19.46  &  56036.46  &  127.01  & $i$ &  20.07 $\pm$ 0.40  \\ 
 2012 Apr 20.37  &  56037.38  &  127.86  & $i$ &  20.42 $\pm$ 0.08  \\ 
 2012 Apr 20.38  &  56037.38  &  127.87  & $g$ &  21.27 $\pm$ 0.11  \\ 
 2012 Apr 22.22  &  56039.22  &  129.58  & $r$ &  20.44 $\pm$ 0.04  \\ 
 2012 May  5.28  &  56052.28  &  141.74  & $r$ &  20.50 $\pm$ 0.25  \\ 
 2012 May  5.28  &  56052.28  &  141.74  & $i$ &  20.16 $\pm$ 0.20  \\ 
 2012 May  6.28  &  56053.28  &  142.66  & $r$ &  20.49 $\pm$ 0.17  \\ 
 2012 May  6.28  &  56053.28  &  142.66  & $i$ &  20.73 $\pm$ 0.20  \\ 
 2012 May  7.25  &  56054.25  &  143.56  & $r$ &  20.36 $\pm$ 0.13  \\ 
 2012 May  7.25  &  56054.25  &  143.56  & $i$ &  20.40 $\pm$ 0.16  \\ 
 2012 May  7.25  &  56054.25  &  143.57  & $g$ &  21.38 $\pm$ 0.39  \\ 
 2012 May  8.26  &  56055.26  &  144.51  & $i$ &  20.58 $\pm$ 0.18  \\ 
 2012 May  8.27  &  56055.27  &  144.51  & $g$ &  21.54 $\pm$ 0.43  \\ 
 2012 May 14.32  &  56061.32  &  150.15  & $i$ &  20.65 $\pm$ 0.07  \\ 
 2012 May 14.32  &  56061.32  &  150.15  & $r$ &  20.56 $\pm$ 0.06  \\ 
 2012 May 14.33  &  56061.33  &  150.16  & $g$ &  21.48 $\pm$ 0.11  \\ 
 2012 May 17.26  &  56064.26  &  152.89  & $i$ &  20.90 $\pm$ 0.10  \\ 
 2012 May 17.27  &  56064.27  &  152.89  & $r$ &  20.66 $\pm$ 0.06  \\ 
 2012 May 17.27  &  56064.27  &  152.89  & $g$ &  21.71 $\pm$ 0.13  \\ 
 2012 May 17.27  &  56064.27  &  152.90  & $i$ &  20.76 $\pm$ 0.09  \\ 
 2012 May 17.28  &  56064.28  &  152.90  & $r$ &  20.65 $\pm$ 0.06  \\ 
 2012 May 17.28  &  56064.28  &  152.90  & $g$ &  21.60 $\pm$ 0.13  \\ 
 2012 May 21.30  &  56068.30  &  156.64  & $i$ &  20.83 $\pm$ 0.11  \\ 
 2012 May 21.30  &  56068.30  &  156.65  & $r$ &  20.70 $\pm$ 0.07  \\ 
 2012 May 21.31  &  56068.31  &  156.65  & $g$ &  21.59 $\pm$ 0.13  \\ 
 2012 May 28.33  &  56075.34  &  163.19  & $i$ &  21.01 $\pm$ 0.16  \\ 
 2012 May 28.34  &  56075.34  &  163.20  & $r$ &  20.62 $\pm$ 0.10  \\ 
 2012 May 28.34  &  56075.34  &  163.20  & $g$ &  21.68 $\pm$ 0.20  \\ 
 2012 Jun  2.25  &  56080.25  &  167.76  & $i$ &  21.08 $\pm$ 0.22  \\ 
 2012 Jun  2.25  &  56080.25  &  167.76  & $r$ &  20.98 $\pm$ 0.15  \\ 
 2012 Jun  5.22  &  56083.21  &  170.53  & $R_{PTF}$ &  20.48 $\pm$ 0.32  \\ 
 2012 Jun  5.25  &  56083.24  &  170.55  & $R_{PTF}$ &  20.59 $\pm$ 0.37  \\ 
 2012 Jun  7.23  &  56085.23  &  172.41  & $R_{PTF}$ &  20.66 $\pm$ 0.25  \\ 
 2012 Jun  7.26  &  56085.26  &  172.43  & $R_{PTF}$ &  20.90 $\pm$ 0.40  \\ 
 2012 Jun  7.30  &  56085.30  &  172.47  & $i$ &  21.16 $\pm$ 0.30  \\ 
\hline
\end{tabular}
\end{table}
\begin{table}
\begin{tabular}{c c c c c } 
\multicolumn{5}{l}{{\bf Table 1} (Continued).} \\
\hline
\hline
Date & MJD$^a$ & Phase$^b$                 &  Filter &  mag$^c$ \\
(UT)   &          &  (days) &           &           \\ 
\hline 
 2012 Jun  7.30  &  56085.30  &  172.47  & $r$ &  20.82 $\pm$ 0.19  \\ 
 2012 Jun  8.24  &  56086.24  &  173.35  & $i$ &  21.00 $\pm$ 0.13  \\ 
 2012 Jun  8.25  &  56086.25  &  173.35  & $r$ &  20.74 $\pm$ 0.09  \\ 
 2012 Jun  8.25  &  56086.25  &  173.35  & $g$ &  21.69 $\pm$ 0.20  \\ 
 2012 Jun  9.20  &  56087.20  &  174.23  & $R_{PTF}$ &  20.58 $\pm$ 0.16  \\ 
 2012 Jun  9.20  &  56087.20  &  174.24  & $R_{PTF}$ &  20.80 $\pm$ 0.20  \\ 
 2012 Jun  9.23  &  56087.23  &  174.26  & $R_{PTF}$ &  20.82 $\pm$ 0.21  \\ 
 2012 Jun  9.26  &  56087.25  &  174.29  & $R_{PTF}$ &  20.68 $\pm$ 0.18  \\ 
 2012 Jun  9.27  &  56087.27  &  174.30  & $R_{PTF}$ &  20.65 $\pm$ 0.19  \\ 
 2012 Jun  9.27  &  56087.27  &  174.31  & $i$ &  20.96 $\pm$ 0.15  \\ 
 2012 Jun  9.28  &  56087.28  &  174.31  & $g$ &  21.86 $\pm$ 0.24  \\ 
 2012 Jun  9.30  &  56087.30  &  174.33  & $R_{PTF}$ &  20.89 $\pm$ 0.37  \\ 
 2012 Jun 10.31  &  56088.31  &  175.27  & $i$ &  21.38 $\pm$ 0.24  \\ 
 2012 Jun 10.31  &  56088.31  &  175.27  & $g$ &  21.92 $\pm$ 0.31  \\ 
 2012 Jun 11.24  &  56089.24  &  176.13  & $R_{PTF}$ &  21.08 $\pm$ 0.26  \\ 
 2012 Jun 11.25  &  56089.25  &  176.14  & $i$ &  20.80 $\pm$ 0.12  \\ 
 2012 Jun 11.25  &  56089.25  &  176.14  & $r$ &  20.84 $\pm$ 0.08  \\ 
 2012 Jun 11.25  &  56089.25  &  176.15  & $g$ &  21.85 $\pm$ 0.17  \\ 
 2012 Jun 11.27  &  56089.27  &  176.16  & $R_{PTF}$ &  20.83 $\pm$ 0.23  \\ 
 2012 Jun 11.30  &  56089.30  &  176.19  & $R_{PTF}$ &  20.88 $\pm$ 0.28  \\ 
 2012 Jun 14.19  &  56092.20  &  178.89  & $R_{PTF}$ &  20.70 $\pm$ 0.16  \\ 
 2012 Jun 14.20  &  56092.20  &  178.89  & $R_{PTF}$ &  20.52 $\pm$ 0.14  \\ 
 2012 Jun 14.23  &  56092.23  &  178.92  & $R_{PTF}$ &  20.84 $\pm$ 0.19  \\ 
 2012 Jun 15.27  &  56093.27  &  179.88  & $i$ &  21.13 $\pm$ 0.14  \\ 
 2012 Jun 16.20  &  56094.20  &  180.75  & $R_{PTF}$ &  20.46 $\pm$ 0.14  \\ 
 2012 Jun 16.21  &  56094.21  &  180.76  & $R_{PTF}$ &  20.61 $\pm$ 0.16  \\ 
 2012 Jun 16.24  &  56094.24  &  180.79  & $R_{PTF}$ &  20.90 $\pm$ 0.21  \\ 
 2012 Jun 18.19  &  56096.20  &  182.61  & $R_{PTF}$ &  20.95 $\pm$ 0.22  \\ 
 2012 Jun 18.20  &  56096.20  &  182.61  & $R_{PTF}$ &  21.18 $\pm$ 0.31  \\ 
 2012 Jun 18.22  &  56096.22  &  182.63  & $R_{PTF}$ &  20.83 $\pm$ 0.21  \\ 
 2012 Jun 20.19  &  56098.20  &  184.47  & $R_{PTF}$ &  21.32 $\pm$ 0.39  \\ 
 2012 Jun 20.20  &  56098.20  &  184.47  & $R_{PTF}$ &  20.84 $\pm$ 0.23  \\ 
 2012 Jun 22.19  &  56100.19  &  186.32  & $R_{PTF}$ &  21.00 $\pm$ 0.27  \\ 
 2012 Jun 22.20  &  56100.20  &  186.33  & $R_{PTF}$ &  20.68 $\pm$ 0.20  \\ 
 2012 Jun 22.23  &  56100.22  &  186.36  & $R_{PTF}$ &  21.19 $\pm$ 0.33  \\ 
 2012 Jun 23.28  &  56101.28  &  187.34  & $i$ &  21.19 $\pm$ 0.29  \\ 
 2012 Jun 23.28  &  56101.28  &  187.34  & $r$ &  20.87 $\pm$ 0.16  \\ 
 2012 Jun 24.22  &  56102.22  &  188.22  & $R_{PTF}$ &  20.42 $\pm$ 0.22  \\ 
 2012 Jun 30.19  &  56108.19  &  193.77  & $R_{PTF}$ &  20.87 $\pm$ 0.39  \\ 
 2012 Jul  4.19  &  56112.19  &  197.50  & $R_{PTF}$ &  20.63 $\pm$ 0.42  \\ 
 2012 Jul  7.18  &  56115.18  &  200.28  & $R_{PTF}$ &  21.10 $\pm$ 0.31  \\ 
 2012 Jul  7.19  &  56115.19  &  200.29  & $R_{PTF}$ &  20.99 $\pm$ 0.26  \\ 
 2012 Jul  7.22  &  56115.22  &  200.32  & $R_{PTF}$ &  20.81 $\pm$ 0.40  \\ 
 2012 Jul  9.19  &  56117.19  &  202.15  & $R_{PTF}$ &  21.10 $\pm$ 0.35  \\ 
 2012 Jul  9.22  &  56117.22  &  202.18  & $R_{PTF}$ &  20.78 $\pm$ 0.28  \\ 
 2012 Jul  9.25  &  56117.25  &  202.20  & $R_{PTF}$ &  20.90 $\pm$ 0.42  \\ 
 2012 Jul 11.19  &  56119.19  &  204.01  & $R_{PTF}$ &  20.06 $\pm$ 0.39  \\ 
 2012 Jul 15.18  &  56123.18  &  207.72  & $R_{PTF}$ &  20.94 $\pm$ 0.33  \\ 
 2012 Jul 15.21  &  56123.21  &  207.75  & $R_{PTF}$ &  20.80 $\pm$ 0.29  \\ 
 2013 Mar 11 & 56362    & 430   &     $B$   &     22.94 $\pm$ 0.19 \\  
 2013 Mar 11 & 56362    & 430    &     $V$   &     22.32 $\pm$ 0.05 \\
 2013 Mar 11 & 56362    & 430    &     $R$   &     22.04 $\pm$ 0.12 \\
 2013 Mar 11 & 56362    & 430    &     $I$   &     22.18 $\pm$ 0.38 \\
\hline
\multicolumn{5}{l}{$^a$Observing epoch ( = JD $-$ 2,400,000.5).} \\
\multicolumn{5}{l}{$^b$In  rest frame, computed from the epoch of estimated}\\ 
\multicolumn{5}{l}{~~ explosion (2011 Dec. 5).}\\
\multicolumn{5}{l}{$^c$Observed apparent magnitudes, with no correction applied.}\\
\multicolumn{5}{l}{~~  The PTF magnitudes ($g,r,i$, and $R_{\rm PTF}$ filters) are in  the AB}\\ 
\multicolumn{5}{l}{~~  system; the VLT FORS2 magnitudes ($BVRI$) are in the Bessell system.}\\
\end{tabular}
\end{table}

\begin{table}
\centering 
\caption{Summary of spectroscopic observations of PTF11rka.}
\label{tab:logspecobs}
\begin{tabular}{ccccc} 
\hline Date  &  MJD$^a$ & Phase$^b$   &  Telescope &  Instrument  \\ 
		(UT) & (days) &                   &                     \\
		\hline
2011  Dec 26  &   55921  & 20  &  Keck-I       & LRIS \\ 
2012  Jan 27  &   55953  & 49   &  KPNO 4\,m &  RC Spectrograph     \\           
2012  Feb 20  &   55977  & 72  &  Keck-I       & LRIS  \\        
2012  Apr 27  &  56044   & 134   &  Keck-I       & LRIS   \\       
2012 May 22  & 56069   & 157   & Keck-I         & LRIS   \\      
2013 Mar 13  & 56364   &  432     & VLT       &  FORS2+300V  \\ 
\hline 
\multicolumn{5}{l}{$^a$Observing epoch ( = JD $-$ 2,400,000.5).}\\
\multicolumn{5}{l}{$^b$In rest-frame, computed from the epoch of estimated explosion (2011 Dec. 5).}\\
\end{tabular}
\end{table}






\bsp	
\label{lastpage}
\end{document}